\documentclass{imammb}

\usepackage{natbib}

\usepackage{graphicx}
\usepackage{amsmath,amsthm,bm,mathrsfs}

\usepackage{rotating}

 \newtheoremstyle{theorem}{6pt}{6pt}{\rm}{}{\sffamily}{ }{ }{}
 \theoremstyle{theorem}

 \newtheoremstyle{algorithm}{6pt}{6pt}{\rm}{}{\sffamily}{ }{ }{}
 \theoremstyle{algorithm}

 \newtheoremstyle{lemma}{6pt}{6pt}{\rm}{}{\sffamily}{ }{ }{}
 \theoremstyle{lemma}

\newtheoremstyle{case}{6pt}{6pt}{\rm}{}{\sffamily}{. }{ }{}
 \theoremstyle{case}

 \newtheoremstyle{statement}{6pt}{6pt}{\rm}{}{\sffamily}{ }{ }{}
\theoremstyle{statement}

 \newtheoremstyle{corollary}{6pt}{6pt}{\rm}{}{\sffamily}{ }{ }{}
 \theoremstyle{corollary}

  \newtheoremstyle{definition}{6pt}{6pt}{\rm}{}{\sffamily}{ }{ }{}
 \theoremstyle{definition}

\newtheoremstyle{example}{6pt}{6pt}{\rm}{}{\sffamily}{ }{ }{}
\theoremstyle{example}

\newtheoremstyle{remark}{6pt}{6pt}{\rm}{}{\sffamily}{ }{ }{}
\theoremstyle{remark}

\newtheoremstyle{approximation}{6pt}{6pt}{\rm}{}{\sffamily}{ }{ }{}
\theoremstyle{approximation}

\newtheoremstyle{scheme}{6pt}{6pt}{\rm}{}{\sffamily}{ }{ }{}
\theoremstyle{scheme}

\newtheoremstyle{Algorithm}{6pt}{6pt}{\rm}{}{\sffamily}{ }{ }{}
\theoremstyle{Algorithm}

\newtheoremstyle{Assumption}{6pt}{6pt}{\rm}{}{\sffamily}{ }{ }{}
\theoremstyle{Assumption}

\newtheoremstyle{proposition}{6pt}{6pt}{\rm}{}{\sffamily}{ }{ }{}
\theoremstyle{proposition}

\newtheoremstyle{hypo}{6pt}{6pt}{\rm}{}{\sffamily}{ }{ }{}
 \theoremstyle{hypo}

  \newtheoremstyle{Step}{6pt}{6pt}{\rm}{}{}{ }{ }{}
 \theoremstyle{Step}

\renewcommand{\theequation}{\thesection.\arabic{equation}}
\numberwithin{equation}{section}

\begin{document}

\title{Optimal conditions for slow 
passive release of heparin-binding 
growth factors from an affinity-based 
delivery system}

\author{ {\sc Tuoi Vo T.N. \& Martin G. Meere}\\[2pt]
School of Mathematics, Statistics and 
Applied Mathematics, \\
National University of Ireland, Galway; \\
University Road, Galway, Ireland.\\[6pt]
\vspace*{6pt}}
\pagestyle{headings}
\markboth{T. VO T.N. AND M.G. MEERE}
{\rm OPTIMAL CONDITIONS FOR SLOW PASSIVE 
RELEASE OF GROWTH FACTORS}
\maketitle

\begin{abstract}
{We consider a mathematical model that
describes the release of heparin-binding
growth factors from an affinity-based 
delivery system. In the delivery system,
heparin binds to a peptide which has been 
covalently cross-linked to a fibrin 
matrix. Growth factor in turn binds to the 
heparin, and growth factor release is 
governed by both binding and diffusion 
mechanisms, the purpose of the binding 
being to slow growth factor release.
The governing mathematical 
model, which in its original formulation 
consists of five partial differential 
equations, is reduced to a system of 
just two equations. We identify the 
governing non-dimensional parameters that 
can be varied to tune the growth factor 
release rate. In particular, we identify a 
parameter regime that ensures slow passive 
release (usually desirable) of at least a 
fraction of the growth factor. It is found 
that slow release is assured if the 
matrix is prepared with the 
concentration of cross-linked peptide 
greatly exceeding the dissociation 
constant of heparin from the peptide, 
and with the concentration of heparin 
greatly exceeding the dissociation 
constant of the growth factor from heparin. 
Also, for the first time, in vitro 
experimental release data is directly 
compared with theoretical release 
profiles generated by the model. We 
propose that the two stage release 
behaviour frequently seen in 
experiments is due to an initial rapid 
out-diffusion of free growth factor 
over a diffusion time scale (typically 
days), followed by a much slower release 
of the bound fraction over a time scale 
depending on both diffusion 
and binding parameters (frequently months).}
{drug delivery; heparin-binding 
growth factor; mathematical model.}

\end{abstract}

\section{Introduction}

\noindent {\em Background}

\vspace{0.3cm}

In verterbrates, the extracellular matrix 
is a complex mixture of carbohydrates, proteins,
and possibly minerals, that surrounds the cells
that form tissues (\cite{Alberts-2002}).
The extracellular matrix helps cells to bind 
together, and regulates a number of cellular 
functions, such as differentiation, 
proliferation, migration, and adhesion. 
The matrix can achieve such regulation via the 
appropriate release of growth factors, for 
which it can act as a depot. Macromolecules 
within the structure of the matrix can bind 
growth factors with high affinity, enabling 
the matrix to serve as a growth factor 
reservoir. In response to changes 
in local physiological conditions (such as
the occurrence of a wound, for example), 
cells may secrete enzymes that can release such 
growth factor depots from the matrix. This 
natural growth factor release mechanism has 
inspired the design of affinity-based drug 
delivery systems that mimic the retentive 
and protective properties of the extracellular 
matrix for growth factor.  In this paper, we 
analyze a mathematical model that describes 
drug release from some such delivery systems, 
make recommendations as to how delivery system 
should be prepared, and, for the first time, 
compare the predictions of the model directly 
with experimental data.   

The extracellular matrix consists predominantly  
of two classes of macromolecules: glycosaminoglycans, 
and fibrous proteins, such as collagen and fibronectin. 
Glycosaminoglycans are polysaccharide polymers that 
typically have a repeating unit consisting of two 
sugars. Heparin is glycosaminoglycan of the matrix 
that is known to bind with a number growth factors 
in vivo via electrostatic interactions, examples 
being basic fibroblast growth factor (bFGF), nerve 
growth factor (NGF), neurotrophin-3 (NT-3), and 
vascular endothelial growth factor (VEGF).

\vspace{0.3cm}

\noindent {\em Affinity-based drug delivery 
systems}

\vspace{0.3cm}

\begin{figure}[h]
\centerline{\includegraphics [width=12cm,height=6cm]
           {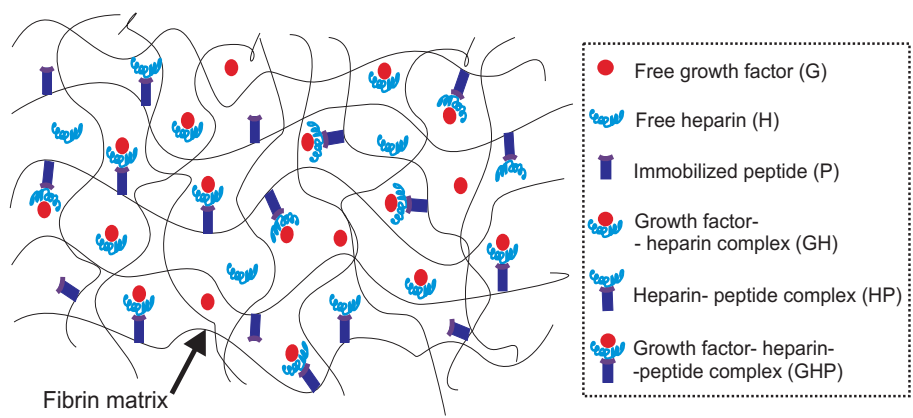} }  
      \vskip -0.2cm
      \caption{\small A schematic representation of
      the fibrin matrix containing all six species
      of the model.}
      \label{fig:fibrin_matrix}     
\end{figure}

\cite{S-Elbert-2000a} have developed a 
growth factor delivery system for wound 
healing that exploits heparin's ability 
to bind electrostatically with growth 
factor. The various elements of their 
system are depicted schematically in 
Figure \ref{fig:fibrin_matrix}.  The 
natural blood clotting  matrix, fibrin, 
was chosen as the base material. Three 
dimensional fibrin hydrogel scaffolds 
were fabricated, into which invading 
cells could infiltrate, and release 
(via enzymatic processes) growth factor 
attached to the fibrin matrix. The 
growth factor attaches to the matrix 
via a bi-domain peptide bound to heparin, 
as we now explain. The peptide contains 
a domain which covalently cross-links 
to the  fibrin matrix. However, the 
peptide also contains a domain that can 
bind to heparin, and such bound heparin 
can in turn bind to heparin-binding 
growth factor. Hence, growth factor 
attachment to the matrix is dependent 
on three distinct interactions, which we 
crudely represent  by: 
(fibrin){\bf --}(peptide){\bf --}(heparin){\bf --}(growth factor).

The peptide is susceptible to cleavage 
by enzymes released from invading cells. 
Cells infiltrate the scaffold, release 
enzymes that degrade the peptide, and 
thereby release growth factor. In these 
systems, it is desirable that the growth
factor be retained by the matrix until 
such time as it is actively released 
by cells. However, there inevitably will 
be some passive release, whereby free 
growth factor diffuses out of the system 
before cells have had the oppurtunity to 
actively release it. Growth factor binding 
to heparin is reversible, and it will not 
be permanently fixed to the matrix even in 
the absence of cells. The passive release 
of growth factor is usually undesirable, 
and in this study we use a mathematical 
model proposed by \cite{S-Elbert-2000a} to 
help identify conditions that minimize it.

\begin{figure}[h]
\centerline{\includegraphics [width=9cm,height=5cm]
           {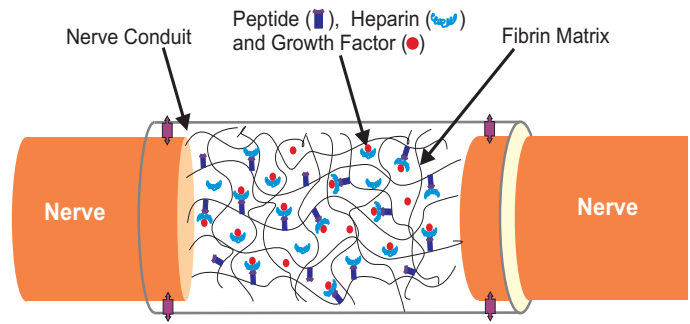} }  
      \vskip -0.2cm
      \caption{\small A schematic representation
      of a nerve guidance conduit containing the 
      growth factor delivery system.}
      \label{fg:nerve_guide_tube}     
\end{figure}

In \cite{S-Elbert-2000a}, the system was 
used to deliver the growth factor bFGF. 
Specifically, they carried out experiments 
in which they placed dorsal root ganglia 
from chickens in fibrin matrices loaded 
with bFGF. The purpose of their experiments 
was to evaluate the effect of the delivery 
system on neurite extension from the dorsal 
root ganglia. Their results demonstrated 
that the delivery system could enhance 
neurite extension by up to about 100\% 
relative to unmodified fibrin. In 
subsequent studies, the system has been 
used to deliver NGF (\cite{S-Elbert-2000b} 
and \cite{Wood-2007, Wood-2009}), 
NT-3 (\cite{Taylor-2004} and 
\cite{Willerth-2008}), glial-derived 
neurotrophic growth factor (GDNF) 
(\cite{Wood-2008,Wood-2009}),
platelet-derived growth factor (PDGF) 
(\cite{Willerth-2008}), and sonic 
hedgehog (\cite{Willerth-2008}).
In \cite{Wood-2009}, a 13 mm gap in
a rat sciatic nerve was repaired using
a silicone nerve guidance conduit 
containing the delivery system loaded 
with GDNF; a schematic representation
of such a conduit is given in 
Figure \ref{fg:nerve_guide_tube}.

\vspace{0.3cm}

\noindent {\em Matrix preparation}

\vspace{0.3cm}

We identify which elements of the 
delivery system may be varied to 
tune growth factor release by 
briefly sketching how the fibrin 
matrices used in the 
above studies were prepared; see
\cite{S-Elbert-2000a} for the
technical details and references.
A fibrinogen solution was prepared 
by mixing the following components
in appropriate quantity to 
achieve desired concentrations: 
plasminogen-free fibrinogen from 
pooled human plasma, calcium ions, 
thrombin, peptide, heparin, and 
growth factor. This polymerization 
mixture was placed in a well of a 
24-well tissue culture plate and 
was incubated under appropriate 
conditions (\cite{S-Elbert-2000a})  
for an hour. It is clear then that the 
concentrations of the peptide, heparin
and growth factor in the polymerization
mixture are easily varied. However, a 
distinction needs to be made between 
the peptide that cross-links to the
fibrin matrix and the peptide that 
remains unbound in the gel since it 
is the cross-linked peptide only that 
forms part of the delivery system. 
In \cite{Sakiyama-1999} 
and \cite{Schense-1999}, a procedure for 
quantifying the fraction of peptide 
that cross-links to the matrix is 
described. If the unbound peptide 
interacts significantly with 
the other free components of the 
system, then account must be taken 
of this in the mathematical modelling, 
and in \cite{Wood-2007} a quite large 
mathematical model incorporating such 
effects is described. However, we do 
not  incorporate free peptide in the 
model considered here for reasons we 
shall discuss in Section 3.

\vspace{0.3cm}

\noindent {\em The system is complex}

\vspace{0.3cm}

\cite{Lin-2006} comments that the 
mathematical model proposed by 
\cite{S-Elbert-2000a} contains a 
relatively large number of parameters 
(we shall see that it has three 
diffusivities and four rate constants), 
and that this complicates their 
modelling approach. Indeed, the model 
in its original formulation does 
contain six differential equations, 
three of which contain diffusion terms. 
A major goal of the current study was 
to simplify the governing mathematical 
model. We shall show that under 
conditions of typical interest, the 
model can be reduced to a standard
system of just two coupled partial 
differential equations governing the 
evolution of the concentrations of 
the total growth factor and total 
heparin. We shall further show that 
under typical conditions, the release 
behaviour is dominated by the values 
of just two non-dimensional parameters: 
the ratio of the concentration of 
cross-linked peptide to the dissociation 
constant of heparin from peptide, and 
the ratio of the initial concentration 
of heparin to the dissociation constant 
of the growth factor from heparin. Lin 
further comments that theoretical profiles
generated by the model had yet to be 
directly compared with experimental data. 
We address this issue in Section 4.     

\section{Mathematical modelling}

\subsection{Model equations}

The mathematical model which we shall 
now describe was first developed 
by \cite{S-Elbert-2000a}.  There are 
six species in all in the model and, 
following \cite{S-Elbert-2000a}, we 
use the following notation:
\begin{eqnarray*}
&&P\mbox{ is an immobile peptide 
covalently fixed to the fibrin matrix;} \\
&&   H\mbox{ is a mobile free heparin 
molecule that can diffuse through the 
fibrin matrix;} \\
&&   G\mbox{ is a mobile free growth 
factor molecule that can diffuse 
through the fibrin matrix;} \\  
&&   GH\mbox{ is a mobile growth 
factor-heparin complex that can 
diffuse through the fibrin matrix;} \\
&&   HP\mbox{ is an immobile 
heparin-peptide complex fixed 
to the fibrin matrix;} \\
&&   GHP\mbox{ is an immobile growth factor-heparin-peptide 
complex fixed to the fibrin matrix.}
\end{eqnarray*}
In Figure \ref{fig:fibrin_matrix}, 
we schematically represent all six 
species in the matrix. The possible 
interactions between these various 
species are described by the 
following four chemical reactions:
\begin{eqnarray}
&& G + H 
   \underset{{k_{\mbox{\tiny r}}}}
   {\overset{{k_{\mbox{\tiny f}}}}
   {\rightleftharpoons}}          GH     \hspace{1.2 cm} 
G + HP 
   \underset{{k_{\mbox{\tiny r}}}}
   {\overset{{k_{\mbox{\tiny f}}}}
   {\rightleftharpoons}}          GHP   \nonumber    \\
&&   \label{chemical_reactions} \\   
&& H + P 
   \underset{{{\mbox{\tiny K}}_{\mbox{\tiny R}}}}
   {\overset{{{\mbox{\tiny K}}_{\mbox{\tiny F}} }}
   {\rightleftharpoons}}          HP      \hspace{1.2 cm}
GH + P 
    \underset{{{\mbox{\tiny K}}_{\mbox{\tiny R}}}}
    {\overset{{{\mbox{\tiny K}}_{\mbox{\tiny F}} }}
    {\rightleftharpoons}}         GHP   \nonumber
\end{eqnarray}
where $k_{f},k_{r},\kappa_{F},\kappa_{R}$ 
are the rate constants as shown. The 
first reaction, for example, 
represents the reversible binding of 
a free growth factor molecule to a 
free heparin molecule, with 
association and dissociation rate 
constants $k_{f}$ and $k_{r}$, 
respectively. The other three 
reactions are similarly interpreted; 
see Figure \ref{fig:fibrin_matrix}. 
It is noteworthy that we have assumed 
that the rate constants for the first 
and second reactions above are the 
same, and similarly for the third 
and fourth reactions. This implies 
that we are assuming that the 
association/dissociation behaviour 
of growth factor for heparin does 
not depend on whether the heparin 
is free or bound to peptide; a 
similar comment applies to the 
binding heparin to peptide.
 
The problems that are considered 
in this paper are one-dimensional, 
and throughout we shall denote the
spatial variable by $x$ and the 
time variable by $t$. Following 
the notation of \cite{S-Elbert-2000a}, 
we denote by 
$c_{\mbox{\tiny G}}(x,t)$ 
the concentration of free growth 
factor $G$ at location $x$ and 
time $t$; the notation for the 
concentrations of the other five 
species follows in an obvious 
fashion. In view
of the chemical reactions 
(\ref{chemical_reactions}), and 
our assumptions regarding the 
mobility of the various species, 
the governing equations for the 
six concentrations take the 
form:
\begin{eqnarray}
&& \frac{\partial c_{\mbox{\tiny G}}}{\partial t} = 
   D_{\mbox{\tiny G}}
   \frac{\partial^2 c_{\mbox{\tiny G}}}
   {\partial x^2}
     - k_f c_{\mbox{\tiny G}} c_{\mbox{\tiny H}} 
     + k_r c_{\mbox{\tiny GH}}
     - k_f c_{\mbox{\tiny G}} c_{\mbox{\tiny HP}} 
     + k_r c_{\mbox{\tiny GHP}},         \nonumber \\
&& \frac{\partial c_{\mbox{\tiny H}}}{\partial t} = 
    D_{\mbox{\tiny H}}
    \frac{\partial^2 c_{\mbox{\tiny H}}}
    {\partial x^2} 
    - k_f c_{\mbox{\tiny G}} c_{\mbox{\tiny H}} 
    + k_r c_{\mbox{\tiny GH}}
    - {\mbox{\tiny K}}_{\mbox{\tiny F}} c_{\mbox{\tiny H}} 
    c_{\mbox{\tiny P}} 
    + {\mbox{\tiny K}}_{\mbox{\tiny R}} c_{\mbox{\tiny HP}}, 
    \nonumber \\
&& \frac{\partial c_{\mbox{\tiny GH}}}{\partial t} = 
    D_{\mbox{\tiny GH}}
    \frac{\partial^2 c_{\mbox{\tiny GH}}}
    {\partial x^2} 
    + k_f c_{\mbox{\tiny G}} c_{\mbox{\tiny H}} 
    - k_r c_{\mbox{\tiny GH}}
    - {\mbox{\tiny K}}_{\mbox{\tiny F}} c_{\mbox{\tiny GH}} 
    c_{\mbox{\tiny P}} 
    + {\mbox{\tiny K}}_{\mbox{\tiny R}} c_{\mbox{\tiny GHP}}, 
    \label{eq:governing_equations}\\
&& \frac{\partial c_{\mbox{\tiny P}}}{\partial t} = 
    - {\mbox{\tiny K}}_{\mbox{\tiny F}} c_{\mbox{\tiny H}} 
    c_{\mbox{\tiny P}} 
    + {\mbox{\tiny K}}_{\mbox{\tiny R}} c_{\mbox{\tiny HP}} 
    - {\mbox{\tiny K}}_{\mbox{\tiny F}} c_{\mbox{\tiny GH}} 
    c_{\mbox{\tiny P}} 
    + {\mbox{\tiny K}}_{\mbox{\tiny R}} c_{\mbox{\tiny GHP}}, 
    \nonumber\\
&& \frac{\partial c_{\mbox{\tiny HP}}}{\partial t} = 
    - k_f c_{\mbox{\tiny G}} c_{\mbox{\tiny HP}} 
    + k_r c_{\mbox{\tiny GHP}}
    + {\mbox{\tiny K}}_{\mbox{\tiny F}} 
      c_{\mbox{\tiny H}} c_{\mbox{\tiny P}} 
    - {\mbox{\tiny K}}_{\mbox{\tiny R}} 
    c_{\mbox{\tiny HP}}, \nonumber\\
&& \frac{\partial c_{\mbox{\tiny GHP}}}{\partial t} = 
    k_f c_{\mbox{\tiny G}} c_{\mbox{\tiny HP}} 
    - k_r c_{\mbox{\tiny GHP}} 
    + {\mbox{\tiny K}}_{\mbox{\tiny F}} 
    c_{\mbox{\tiny GH}} c_{\mbox{\tiny P}} 
    - {\mbox{\tiny K}}_{\mbox{\tiny R}} 
    c_{\mbox{\tiny GHP}}, \nonumber
\end{eqnarray}
where $D_{\mbox{\tiny G}}$, 
$D_{\mbox{\tiny H}}$ and 
$D_{\mbox{\tiny GH}}$ are the 
diffusivities for the free growth 
factor, free heparin, and free growth 
factor-heparin complex, respectively; 
the species $P$, $HP$ and $GHP$ are 
taken to be immobile since the 
peptide is assumed to be covalently 
fixed to the fibrin matrix, 
and consequently the equations 
for their concentrations do not 
contain diffusion terms.

\subsection{Boundary and initial 
conditions}
\begin{figure}[h]
\centerline{\includegraphics [width=7cm,height=5cm]
           {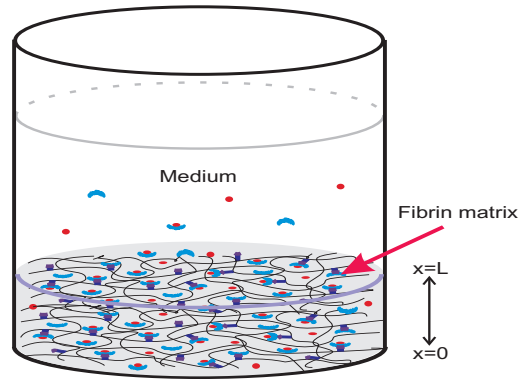} }  
      \vskip -0.2cm
      \caption{\small A schematic depiction of 
      the experimental setup for the release of
      growth factor from the fibrin matrix.}
      \label{fig:experimental_setup}     
\end{figure}
We choose simple boundary conditions 
for (\ref{eq:governing_equations}) 
that allow direct comparison with 
available experimental and theoretical 
results for growth factor release from 
the delivery system. We suppose that 
the fibrin matrix occupies 
$0\leq x\leq L$, with  $x=0$ giving 
the location of a container wall 
through which the growth factor cannot 
penetrate, and $x=L$ denoting the 
interface between the matrix and 
an external medium into which growth 
factor releases; see 
Figure \ref{fig:experimental_setup}. 
At the container wall, we impose 
no-flux conditions for the mobile 
species, so that:  
\begin{equation}
\frac{\partial c_{\mbox{\tiny G}}}{\partial x} 
 = 0,\hspace{0.2cm}
\frac{\partial c_{\mbox{\tiny H}}}{\partial x}
 = 0,\hspace{0.2cm}
\frac{\partial c_{\mbox{\tiny GH}}}{\partial x}
 = 0\hspace{0.4cm}\mbox{ on }x=0.
\label{eq:containerbcs}
\end{equation}
At the interface between the matrix 
and the external medium, we impose 
perfect sink conditions for the 
mobile species: 
\begin{equation}
c_{\mbox{\tiny G}}=0,\hspace{0.2cm}
c_{\mbox{\tiny H}}=0,\hspace{0.2cm}
c_{\mbox{\tiny GH}}=0\hspace{0.4cm}\mbox{ on }x=L.
\label{eq:interfacebcs}
\end{equation}
The boundary conditions 
(\ref{eq:containerbcs}), 
(\ref{eq:interfacebcs}) could also 
model in-vitro growth factor release 
from a nerve guide 
tube (\cite{S-Elbert-2000a}) 
occupying $-L\leq x \leq L$, 
with $x=\pm L$ giving the location of 
the ends of the tube; see 
Figure \ref{fg:nerve_guide_tube}. In 
this context, (\ref{eq:containerbcs}) 
are interpreted as symmetry conditions 
on the centre-line of the tube.

Equations (\ref{eq:governing_equations}) 
are solved subject to the following 
initial conditions:
\begin{equation}
c_{\mbox{\tiny G}} = 
c_{\mbox{\tiny G}}^{\mbox{\tiny 0}}, \hspace{0.2cm}   
c_{\mbox{\tiny H}} = 
c_{\mbox{\tiny H}}^{\mbox{\tiny 0}}, \hspace{0.2cm}
c_{\mbox{\tiny GH}} = 0, \hspace{0.2cm}
c_{\mbox{\tiny P}} = 
c_{\mbox{\tiny P}}^{\mbox{\tiny 0}}, \hspace{0.2cm}   
c_{\mbox{\tiny HP}} = 0, \hspace{0.2cm}
c_{\mbox{\tiny GHP}} = 0 \hspace{0.4cm}
\mbox{ at }t=0,
\label{eq:initial_conditions}
\end{equation}
where 
$c_{\mbox{\tiny G}}^{\mbox{\tiny 0}}, 
c_{\mbox{\tiny H}}^{\mbox{\tiny 0}},  
c_{\mbox{\tiny P}}^{\mbox{\tiny 0}}$ 
denote the initial concentrations of 
growth factor, heparin, and peptide, 
respectively, in the polymerization
mixture. In our modelling, we make 
the simplifying assumption that all 
of the peptide in the  
polymerization mixture crosslinks
covalently to the fibrin matrix. 
This is not likely to occur in 
practice, but we shall show in 
Section 3 how our results may be 
modified to take account of the 
presence of free peptide in the 
matrix. It should also be noted 
that free peptide will typically 
clear the system over a period 
of a few days. We conclude 
this section by emphasising 
that the key results of this 
paper are quite general for 
the model under typical 
conditions, and are not strongly 
dependent on the particular 
choice of boundary and initial 
conditions made.

The mathematical model is now 
complete, and consists of 
equations 
(\ref{eq:governing_equations}),
(\ref{eq:containerbcs}),
(\ref{eq:interfacebcs}),
(\ref{eq:initial_conditions}).

\subsection{Model reduction}

We now show how the model may 
frequently be reduced to a coupled 
pair of partial differential 
equations.

\subsubsection{Non-dimensionalisation} 

Before giving the non-dimensionalisation, 
we first note that 
(\ref{eq:governing_equations}) may be 
written in the following equivalent form: 
\begin{eqnarray}
&& 
 \frac{\partial }{\partial t}({c_{\mbox{\tiny G}}
   +  c_{\mbox{\tiny GH}} + c_{\mbox{\tiny GHP}}}) 
   =  D_{\mbox{\tiny G}}  
      \frac{\partial^2 c_{\mbox{\tiny G}}}
      {\partial x^2}
   +  D_{\mbox{\tiny GH}}
      \frac{\partial^2 c_{\mbox{\tiny GH}}}
      {\partial x^2},   \nonumber \\
&&  \frac{\partial}{\partial t}\left(
    c_{\mbox{\tiny H}} + c_{\mbox{\tiny GH}} 
 +  c_{\mbox{\tiny HP}} +  c_{\mbox{\tiny GHP}} 
   \right) = 
   D_{\mbox{\tiny H}}  
   \frac{\partial^2 c_{\mbox{\tiny H}}}
   {\partial x^2} 
 + D_{\mbox{\tiny GH}} 
   \frac{\partial^2 c_{\mbox{\tiny GH}}}
   {\partial x^2},   \nonumber \\
&&   \frac{\partial }{\partial t}({c_{\mbox{\tiny GH}} 
 +   c_{\mbox{\tiny GHP}}}) = 
     D_{\mbox{\tiny GH}}      
     \frac{\partial^2 c_{\mbox{\tiny GH}}}
     {\partial x^2}
 +   k_f  c_{\mbox{\tiny G}}(c_{\mbox{\tiny H}} 
 +   c_{\mbox{\tiny HP}}) - k_r (c_{\mbox{\tiny GH}} 
 +   c_{\mbox{\tiny GHP}}),
        \label{eq:rearranged_governing_equations}  \\
&& \frac{\partial }{\partial t} \left(
    c_{\mbox{\tiny HP}} 
 +  c_{\mbox{\tiny GHP}} \right) = 
    {\mbox{\tiny K}}_{\mbox{\tiny F}} 
    c_{\mbox{\tiny P}} 
    \left( c_{\mbox{\tiny H}} 
 +  c_{\mbox{\tiny GH}} \right) 
 - {\mbox{\tiny K}}_{\mbox{\tiny R}}   
   \left( c_{\mbox{\tiny HP}} 
 + c_{\mbox{\tiny GHP}} \right),
          \nonumber  \\
&& \frac{\partial c_{\mbox{\tiny GHP}}}{\partial t} = 
     k_f  c_{\mbox{\tiny G}} c_{\mbox{\tiny HP}}
 +   {\mbox{\tiny K}}_{\mbox{\tiny F}}  
     c_{\mbox{\tiny GH}} c_{\mbox{\tiny P}} 
 -   (k_r + {\mbox{\tiny K}}_{\mbox{\tiny R}}) 
     c_{\mbox{\tiny GHP}},
          \nonumber  \\
&&  c_{\mbox{\tiny P}}+c_{\mbox{\tiny HP}} 
 +  c_{\mbox{\tiny GHP}}
 =  c_{\mbox{\tiny P}}^{\mbox{\tiny 0}},
          \nonumber
\end{eqnarray}
where, for example, equation
(\ref{eq:rearranged_governing_equations})$_{6}$
is obtained by forming 
(\ref{eq:governing_equations})$_{4}$+
(\ref{eq:governing_equations})$_{5}$+
(\ref{eq:governing_equations})$_{6}$.  
We denote the total concentrations of growth 
factor and heparin in the matrix by 
$c_{\mbox{\tiny G}}^{\mbox{\tiny T}}$ and 
$c_{\mbox{\tiny H}}^{\mbox{\tiny T}}$, 
respectively, so that: 
\begin{equation}
 c_{\mbox{\tiny G}}^{\mbox{\tiny T}} = 
 c_{\mbox{\tiny G}} + c_{\mbox{\tiny GH}} 
   + c_{\mbox{\tiny GHP}}, \hspace{0.2cm}
 c_{\mbox{\tiny H}}^{\mbox{\tiny T}} = 
   c_{\mbox{\tiny H}}+ c_{\mbox{\tiny GH}} 
   + c_{\mbox{\tiny HP}} + c_{\mbox{\tiny GHP}}. 
\label{eq:growth_factor_heparin_totals}
\end{equation}
Equations 
(\ref{eq:rearranged_governing_equations})$_{1}$
and 
(\ref{eq:rearranged_governing_equations})$_{2}$ 
give the evolution equations for the total growth 
factor and heparin, and may be written in 
conservation form as:
\begin{equation}
\frac{\partial c_{\mbox{\tiny G}}^{\mbox{\tiny T}}}
  {\partial t} + 
  \frac{\partial j_{\mbox{\tiny G}}^{\mbox{\tiny T}}}
  {\partial x} = 0, \hspace{0.3cm}
\frac{\partial c_{\mbox{\tiny H}}^{\mbox{\tiny T}}}
  {\partial t} + 
  \frac{\partial j_{\mbox{\tiny H}}^{\mbox{\tiny T}}}
  {\partial x} = 0,
\label{eq:conservation_of_growth_factor_and_heparin}
\end{equation}
where
\begin{equation}
j_{\mbox{\tiny G}}^{\mbox{\tiny T}} =
      -D_{\mbox{\tiny G}}  
      \frac{\partial c_{\mbox{\tiny G}}}
      {\partial x}
      -D_{\mbox{\tiny GH}}
      \frac{\partial c_{\mbox{\tiny GH}}}
      {\partial x},  \hspace{0.3cm}
j_{\mbox{\tiny H}}^{\mbox{\tiny T}} =
      -D_{\mbox{\tiny H}}  
      \frac{\partial c_{\mbox{\tiny H}}}
      {\partial x} 
      - D_{\mbox{\tiny GH}} 
      \frac{\partial c_{\mbox{\tiny GH}}}
      {\partial x},
\end{equation}
give the total flux of growth factor and 
heparin, respectively.

We introduce non-dimensional variables as 
follows: 
\begin{eqnarray*}
&& \bar{x}=\frac{x}{L}, \hspace{0.2cm}
\bar{t}=\frac{t}
{(L^{2}/D_{\mbox{\tiny GH}})}, \hspace{0.2cm}
\bar{c}_{\mbox{\tiny G}}=\frac{c_{\mbox{\tiny G}}}
{c_{\mbox{\tiny G}}^{\mbox{\tiny 0}}}, 
\hspace{0.2cm} 
\bar{c}_{\mbox{\tiny H}}=\frac{c_{\mbox{\tiny H}}}
{c_{\mbox{\tiny H}}^{\mbox{\tiny 0}}}, 
\hspace{0.2cm} 
\bar{c}_{\mbox{\tiny P}}=\frac{c_{\mbox{\tiny P}}}
  {c_{\mbox{\tiny P}}^{\mbox{\tiny 0}}}, 
\hspace{0.2cm}
\bar{c}_{\mbox{\tiny GH}} = 
  \frac{c_{\mbox{\tiny GH}}}
  {c_{\mbox{\tiny G}}^{\mbox{\tiny 0}}}, \\
&& \bar{c}_{\mbox{\tiny GHP}} = 
  \frac{c_{\mbox{\tiny GHP}}}
  {c_{\mbox{\tiny G}}^{\mbox{\tiny 0}}}, 
\hspace{0.2cm}
\bar{c}_{\mbox{\tiny HP}} = 
\frac{c_{\mbox{\tiny HP}}}
{c_{\mbox{\tiny H}}^{\mbox{\tiny 0}}}, 
\hspace{0.2cm}
\bar{c}_{\mbox{\tiny G}}^{\mbox{\tiny T}} = 
\frac{c_{\mbox{\tiny G}}^{\mbox{\tiny T}}}
{c_{\mbox{\tiny G}}^{\mbox{\tiny 0}}}, 
\hspace{0.2cm}
\bar{c}_{\mbox{\tiny H}}^{\mbox{\tiny T}}=
\frac{c_{\mbox{\tiny H}}^{\mbox{\tiny T}}}
{c_{\mbox{\tiny H}}^{\mbox{\tiny 0}}}, 
\hspace{0.2cm}
\bar{j}_{\mbox{\tiny G}}^{\mbox{\tiny T}}=
\frac{j_{\mbox{\tiny G}}^{\mbox{\tiny T}}}
{(D_{\mbox{\tiny GH}} 
c_{\mbox{\tiny G}}^{\mbox{\tiny 0}} / L ) }, 
\hspace{0.2cm} 
\bar{j}_{\mbox{\tiny H}}^{\mbox{\tiny T}}=
\frac{j_{\mbox{\tiny H}}^{\mbox{\tiny T}}}
{(D_{\mbox{\tiny GH}} 
c_{\mbox{\tiny H}}^{\mbox{\tiny 0}}/L)},
\end{eqnarray*}
to obtain the following non-dimensional 
form for the governing initial boundary 
value problem (upon dropping over-bars):
\begin{eqnarray}
&& 
 \frac{\partial }{\partial t}({c_{\mbox{\tiny G}}
   +  c_{\mbox{\tiny GH}} + c_{\mbox{\tiny GHP}}}) 
   =  D_{\mbox{\tiny G}}^{*}  
      \frac{\partial^2 c_{\mbox{\tiny G}}}
      {\partial x^2}
   +  \frac{\partial^2 c_{\mbox{\tiny GH}}}
   {\partial x^2},   \nonumber \\
&&  \frac{\partial}{\partial t}\left(
    c_{\mbox{\tiny H}} 
 +  c_{\mbox{\tiny HP}} 
 +  (c_{\mbox{\tiny GH}} 
 +  c_{\mbox{\tiny GHP}})/
    \eta_{\mbox{\tiny H/G}} 
    \right) = 
    D_{\mbox{\tiny H}}^{*}   
    \frac{\partial^2 c_{\mbox{\tiny H}}}
    {\partial x^2} 
 +  \frac{1}{\eta_{\mbox{\tiny H/G}}}
    \frac{\partial^2 c_{\mbox{\tiny GH}}}
    {\partial x^2},  \nonumber \\
&&   \delta_{\mbox{\tiny G}} 
     \frac{\partial }{\partial t}
     ({c_{\mbox{\tiny GH}} + c_{\mbox{\tiny GHP}}}) 
 =   \delta_{\mbox{\tiny G}}
     \frac{\partial^2 c_{\mbox{\tiny GH}}}
     {\partial x^2}
 +   K_{b\mbox{\tiny H}}
     c_{\mbox{\tiny G}}(
     c_{\mbox{\tiny H}} 
 +   c_{\mbox{\tiny HP}}) - (c_{\mbox{\tiny GH}} 
 +   c_{\mbox{\tiny GHP}}),
        \nonumber  \\
&& \delta_{\mbox{\tiny H}} \frac{\partial }
    {\partial t} 
    \left( \eta_{\mbox{\tiny H/G}} 
    c_{\mbox{\tiny HP}} 
 +  c_{\mbox{\tiny GHP}} \right) = 
    K_{b\mbox{\tiny P}} c_{\mbox{\tiny P}} 
    \left( \eta_{\mbox{\tiny H/G}} 
    c_{\mbox{\tiny H}} 
 +  c_{\mbox{\tiny GH}} \right) 
 -  \left( \eta_{\mbox{\tiny H/G}}
 c_{\mbox{\tiny HP}} 
 +  c_{\mbox{\tiny GHP}} \right),
  \label{eq:nondimensional_governing_equations}  \\
&&  \delta_{\mbox{\tiny G}}\delta_{\mbox{\tiny H}}
    \frac{\partial c_{\mbox{\tiny GHP}}}{\partial t} 
 =  \delta_{\mbox{\tiny H}} ( K_{b\mbox{\tiny H}}
     c_{\mbox{\tiny G}} c_{\mbox{\tiny HP}}
 -   c_{\mbox{\tiny GHP}} )
 +   \delta_{\mbox{\tiny G}} ( K_{b\mbox{\tiny P}} 
     c_{\mbox{\tiny GH}} c_{\mbox{\tiny P}}
 -   c_{\mbox{\tiny GHP}} ),
          \nonumber  \\
&&  c_{\mbox{\tiny P}}
 +  \eta_{\mbox{\tiny H/P}} c_{\mbox{\tiny HP}}
 +  \frac{\eta_{\mbox{\tiny H/P}}}{
    \eta_{\mbox{\tiny H/G}}} c_{\mbox{\tiny GHP}}
 =   1, \nonumber \\
&&\frac{\partial c_{\mbox{\tiny G}}}{\partial x}
 = 0,  \hspace{0.2cm}
\frac{\partial c_{\mbox{\tiny H}}}{\partial x}
 = 0,  \hspace{0.2cm}
\frac{\partial c_{\mbox{\tiny GH}}}{\partial x}
 = 0  \hspace{0.2cm}
   \mbox{ on }x=0, \nonumber \\
&& c_{\mbox{\tiny G}}=0,\hspace{0.2cm}
c_{\mbox{\tiny H}}=0,\hspace{0.2cm}
c_{\mbox{\tiny GH}}=0\hspace{0.2cm}\mbox{ on }x=1, \nonumber \\
&& c_{\mbox{\tiny G}} = 1,\hspace{0.2cm}   
c_{\mbox{\tiny H}} = 1,\hspace{0.2cm}
c_{\mbox{\tiny GH}} = 0,\hspace{0.2cm}
c_{\mbox{\tiny P}} = 1,\hspace{0.2cm}   
c_{\mbox{\tiny HP}} = 0,\hspace{0.2cm}
c_{\mbox{\tiny GHP}} = 0\hspace{0.2cm}
\mbox{ at }t=0,   \nonumber
\end{eqnarray}
where:
\begin{eqnarray}
&& D_{\mbox{\tiny G}}^* = 
 \frac{D_{\mbox{\tiny G}}}{D_{\mbox{\tiny GH}}}, 
 \hspace{0.2cm} 
 D_{\mbox{\tiny H}}^* = 
 \frac{D_{\mbox{\tiny H}}}{D_{\mbox{\tiny GH}}}, 
 \hspace{0.2cm}
 \eta_{\mbox{\tiny H/G}} = 
 \frac{c_{\mbox{\tiny H}}^{\mbox{\tiny 0}}}
 {c_{\mbox{\tiny G}}^{\mbox{\tiny 0}}}, 
 \hspace{0.2cm}
 \eta_{\mbox{\tiny H/P}} = 
 \frac{c_{\mbox{\tiny H}}^{\mbox{\tiny 0}}}
 {c_{\mbox{\tiny P}}^{\mbox{\tiny 0}}},  
   \nonumber \\
&& \label{eq:dimensionless_parameters} \\
&& \delta_{\mbox{\tiny G}} = \frac{D_{\mbox{\tiny GH}}}
   {k_r L^{2}}, \hspace{0.2cm} 
   \delta_{\mbox{\tiny H}} = 
   \frac{D_{\mbox{\tiny GH}}}
   {{\mbox{\tiny K}}_{\mbox{\tiny R}}L^{2}}, \hspace{0.2cm}
   K_{b{\mbox{\tiny H}}} =  
   \frac{k_f c_{\mbox{\tiny H}}^{\mbox{\tiny 0}}}{k_r}, 
   \hspace{0.2cm}
   K_{b{\mbox{\tiny P}}} = 
   \frac{{\mbox{\tiny K}}_{\mbox{\tiny F}} 
   c_{\mbox{\tiny P}}^{\mbox{\tiny 0}}}
   {{\mbox{\tiny K}}_{\mbox{\tiny R}}}, \nonumber
\end{eqnarray}
are the governing non-dimensional parameters. 
The quantities $K_{b{\mbox{\tiny H}}},
K_{b{\mbox{\tiny P}}}$ give a non-dimensional 
measure of the strength of retention of  
growth factor by the heparin, and of heparin 
by the peptide, respectively. We denote by
\[ 
K_{\mbox{\tiny G-H}}^{\mbox{\tiny D}} = 
\frac{k_{r}}{k_{f}}, \hspace{0.2cm}
K_{\mbox{\tiny H-P}}^{\mbox{\tiny D}} = 
   \frac{{\mbox{\tiny K}}_{\mbox{\tiny R}}}
   {{\mbox{\tiny K}}_{\mbox{\tiny F}}},
\]
the dissociation constants of growth 
factor from heparin, and of heparin 
from peptide, respectively, so that:
\begin{equation} 
K_{b{\mbox{\tiny H}}} = 
\frac{c_{\mbox{\tiny H}}^{\mbox{\tiny 0}}}
{K_{\mbox{\tiny G-H}}^{\mbox{\tiny D}}}, 
\hspace{0.2cm}
K_{b{\mbox{\tiny P}}} = 
\frac{c_{\mbox{\tiny P}}^{\mbox{\tiny 0}}}
{K_{\mbox{\tiny H-P}}^{\mbox{\tiny D}}}. 
\label{eq:important_ratios}
\end{equation}
It is noteworthy that 
$K_{b{\mbox{\tiny H}}}$ involves 
both the concentration of available 
binding sites for the growth factor 
and the dissociation constant of 
growth factor from heparin. The 
case $K_{b{\mbox{\tiny H}}}\gg 1$ 
corresponds to the growth factor 
being {\em strongly} retained 
by the heparin. Conversely, 
$K_{b{\mbox{\tiny H}}}\ll 1$ 
corresponds to weak retention 
of growth factor by the heparin. 
The parameter 
$K_{b{\mbox{\tiny P}}}$ is 
similarly interpreted in the 
context of heparin retention 
by the peptide. 

The non-dimensional form for the 
total growth factor and heparin 
and their fluxes are given by: 
\begin{eqnarray}
&& c_{\mbox{\tiny G}}^{\mbox{\tiny T}} = 
    c_{\mbox{\tiny G}} 
 +  c_{\mbox{\tiny GH}} 
 +  c_{\mbox{\tiny GHP}}, \hspace{0.5cm}
c_{\mbox{\tiny H}}^{\mbox{\tiny T}} =  
   c_{\mbox{\tiny H}}
 + c_{\mbox{\tiny HP}} 
 + (c_{\mbox{\tiny GH}} 
 + c_{\mbox{\tiny GHP}})/
 \eta_{\mbox{\tiny H/G}}, \nonumber \\
&&  \label{eq:nondimensional_totals}  \\
&& j_{\mbox{\tiny G}}^{\mbox{\tiny T}} =
      -D_{\mbox{\tiny G}}^{*} 
      \frac{\partial c_{\mbox{\tiny G}}}{\partial x}
      -\frac{\partial c_{\mbox{\tiny GH}}}{\partial x}, 
      \hspace{0.5cm}
j_{\mbox{\tiny H}}^{\mbox{\tiny T}} =
      -D_{\mbox{\tiny H}}^{*}  
      \frac{\partial c_{\mbox{\tiny H}}}{\partial x} 
      -\frac{1}{\eta_{\mbox{\tiny H/G}}}
      \frac{\partial c_{\mbox{\tiny GH}}}{\partial x}. 
      \nonumber 
\end{eqnarray}

\subsubsection{Parameter values}                                 

In \cite{S-Elbert-2000a, Taylor-2004, 
Willerth-2006, Willerth-2008} and 
\cite{Wood-2007, Wood-2008}, 
the fibrin gels were prepared by placing 
400 $\mu$l of polymerization mixture in 
the wells of a 24-well plate. The diameter 
of each well in such a plate is 1.56 cm, 
from which it follows that the thickness 
of the gels was $L\approx 0.2$ cm. In 
\cite{S-Elbert-2000a}, the heparin 
diffusivity was taken to be 
$D_{\mbox{\tiny H}}=3.13\times 10^{-5}$ 
cm$^{2}$min$^{-1}$, and for bFGF, the 
diffusivities 
$D_{\mbox{\tiny G}}=
6.0\times 10^{-5}$ cm$^{2}$min$^{-1}$
and 
$D_{\mbox{\tiny GH}}=
1.0\times 10^{-5}$ cm$^{2}$min$^{-1}$
were used. These values, which were based 
on the work of \cite{Saltzman-1994} 
and \cite{Gaigalas-1995}, all 
have order of magnitude  
$10^{-5}$ cm$^{2}$min$^{-1}$. 
In \cite{Taylor-2004}, where the growth 
factor being considered was NT-3, the 
diffusivities used were again of order 
$10^{-5}$ cm$^{2}$min$^{-1}$. Taking 
$D=3.0\times 10^{-5}$ cm$^{2}$min$^{-1}$ 
as a representative diffusivity for a 
free species in the matrix and $L=0.2$ cm, 
we calculate a typical diffusion time 
scale for the system to be $L^{2}/D\approx 1$ 
day. Hence, in  a release experiment 
where the diffusivities are of order 
$10^{-5}$ cm$^{2}$min$^{-1}$, 
we would typically expect the unbound 
components to clear the system over a 
period of some days, and this is 
consistent with the experimental 
results of 
\cite{Taylor-2004} and 
\cite{Wood-2007, Wood-2008}.

There are also time scales 
associated with the rate constants 
for the chemical reactions 
(\ref{chemical_reactions}), namely,
$1/k_{r}$, 
$1/(k_{f}c_{\mbox{\tiny H}}^{\mbox{\tiny 0}})$,
$1/\mbox{\tiny K}_{\mbox{\tiny R}}$, and
$1/({\mbox{\tiny K}}_{\mbox{\tiny F}}
c_{\mbox{\tiny P}}^{\mbox{\tiny 0}})$. 
In \cite{S-Elbert-2000a} 
and \cite{Taylor-2004}, the 
values of the rate constants for the 
binding of heparin to the peptide were 
taken to be 
$\mbox{\tiny K}_{\mbox{\tiny R}}
\approx 80$ min$^{-1}$ and
$\mbox{\tiny K}_{\mbox{\tiny F}}
\approx 10^{9}$ M$^{-1}$ min$^{-1}$, 
and 
$c_{\mbox{\tiny P}}^{\mbox{\tiny 0}}$
was of order $10^{-4}$ M. 
For these values, we find that 
$1/\mbox{\tiny K}_{\mbox{\tiny R}}\approx 1$ s,
$1/({\mbox{\tiny K}}_{\mbox{\tiny F}}
c_{\mbox{\tiny P}}^{\mbox{\tiny 0}})
= 10^{-5}$ s, and we note that these 
times are tiny compared to the typical 
time scales associated with diffusion 
(days), and furthermore, would remain 
so even if we made 
$c_{\mbox{\tiny P}}^{\mbox{\tiny 0}}$
orders of magnitude smaller. For 
the  binding of growth factor to 
heparin, the rate constants will depend 
on the nature of the growth factor, and 
data is unfortunately frequently 
lacking. For bFGF, \cite{S-Elbert-2000a} 
use the values 
$k_{r}\approx 1$ min$^{-1}$,
$k_{f}\approx 10^{8}$ M$^{-1}$ min$^{-1}$, 
and take 
$c_{\mbox{\tiny H}}^{\mbox{\tiny 0}}\approx
6\times 10^{-5}$ M. For these values,
$1/k_{r}=1$ min and 
$1/(k_{f}c_{\mbox{\tiny H}}^{\mbox{\tiny 0}})
\approx 10^{-2}$ s, and these times
are also small compared to the diffusion
time scales.

For NT-3, the $k_{f}$ and $k_{r}$ values
are unknown, but \cite{Taylor-2004} gives 
the approximation
$K_{D}^{\mbox{\tiny G-H}}\approx 10^{-6}$ M
for the dissociation constant, which would
imply that NT-3 has a low affinity for the 
heparin binding site. By contrast, bFGF has 
a relatively high affinity for the heparin 
binding site, with dissociation constant 
$K_{D}^{\mbox{\tiny G-H}}\approx 10^{-8}$ M. 
However, we shall show in this paper that, 
provided the governing mathematical model 
is appropriate, slow passive growth factor 
release is achieveable even for low affinity 
binding of growth factor to heparin.

\subsubsection{Reduction to a pair of coupled 
partial differential equations}

We conclude from the remarks above 
that for many systems of practical 
interest, the time scales for the 
association and dissociation rates 
in the chemical reactions 
(\ref{chemical_reactions}) are much 
shorter than the diffusion time 
scales, so that we frequently have:
\[ 
\frac{L^{2}}{D_{\mbox{\tiny G}}},\hspace{0.2cm}
\frac{L^{2}}{D_{\mbox{\tiny H}}},\hspace{0.2cm}
\frac{L^{2}}{D_{\mbox{\tiny GH}}} \gg 
\frac{1}{k_{r}}, \hspace{0.2cm}
\frac{1}{k_{f}c_{\mbox{\tiny H}}^{\mbox{\tiny 0}}},
\hspace{0.2cm}
\frac{1}{{\mbox{\tiny K}}_{\mbox{\tiny R}}}, 
\hspace{0.2cm}
\frac{1}{{\mbox{\tiny K}}_{\mbox{\tiny F}}
c_{\mbox{\tiny P}}^{\mbox{\tiny 0}}}.
\] 
In such cases, diffusion is rate limiting
since it is the slowest process. We restrict 
our attention to such systems in the current 
analysis. In terms of the dimensionless 
parameters (\ref{eq:dimensionless_parameters}), 
the conditions above imply that
$\delta{\mbox{\tiny G}} \ll 
\min(K_{b{\mbox{\tiny H}}},1)$ and  
$\delta{\mbox{\tiny H}} \ll 
\min(K_{b{\mbox{\tiny P}}},1)$, and so the 
differential equations 
(\ref{eq:nondimensional_governing_equations})$_{3}$,
(\ref{eq:nondimensional_governing_equations})$_{4}$
and
(\ref{eq:nondimensional_governing_equations})$_{5}$
are replaced by the algebraic expressions:
\begin{eqnarray}
&& K_{b\mbox{\tiny H}} 
     c_{\mbox{\tiny G}}(
     c_{\mbox{\tiny H}} 
 +   c_{\mbox{\tiny HP}}) = c_{\mbox{\tiny GH}} 
 +   c_{\mbox{\tiny GHP}}, \hspace{0.2cm}
K_{b\mbox{\tiny P}} c_{\mbox{\tiny P}} 
    \left( \eta_{\mbox{\tiny H/G}} 
    c_{\mbox{\tiny H}} 
 +  c_{\mbox{\tiny GH}} \right) 
 =   \eta_{\mbox{\tiny H/G}} c_{\mbox{\tiny HP}} 
 +  c_{\mbox{\tiny GHP}},
  \label{eq:equilibrium_equations} \\
&& \theta ( K_{b\mbox{\tiny H}}
     c_{\mbox{\tiny G}} c_{\mbox{\tiny HP}}
 -   c_{\mbox{\tiny GHP}} )
 +     K_{b\mbox{\tiny P}} 
     c_{\mbox{\tiny GH}} c_{\mbox{\tiny P}}
 -   c_{\mbox{\tiny GHP}} = 0,\nonumber    
\end{eqnarray}
where $\theta = \delta{\mbox{\tiny H}} /
\delta{\mbox{\tiny G}}=k_{r}/
{\mbox{\tiny K}}_{\mbox{\tiny R}}$. 
The first two equations in 
(\ref{eq:equilibrium_equations})
correspond to the equilibrium forms
for the binding of growth factor to heparin, 
and of heparin to peptide, respectively. 

We solve the six algebraic expressions 
(\ref{eq:nondimensional_governing_equations})$_{6}$,
(\ref{eq:nondimensional_totals})$_{1}$,
(\ref{eq:equilibrium_equations})
for the concentrations of the six species
$c_{\mbox{\tiny G}}$, $c_{\mbox{\tiny H}}$,
$c_{\mbox{\tiny GH}}$, $c_{\mbox{\tiny P}}$,
$c_{\mbox{\tiny HP}}$, $c_{\mbox{\tiny GHP}}$
in terms of the total concentration of
growth factor, 
$c_{\mbox{\tiny G}}^{\mbox{\tiny T}}$,
and heparin, 
$c_{\mbox{\tiny H}}^{\mbox{\tiny T}}$, 
to obtain expressions of the form:
\begin{eqnarray*}
&& c_{\mbox{\tiny G}} = c_{\mbox{\tiny G}}(
   c_{\mbox{\tiny G}}^{\mbox{\tiny T}},
   c_{\mbox{\tiny H}}^{\mbox{\tiny T}}),
\hspace{0.2cm}
c_{\mbox{\tiny P}} = c_{\mbox{\tiny P}}(
    c_{\mbox{\tiny H}}^{\mbox{\tiny T}}), 
\hspace{0.2cm}
c_{\mbox{\tiny GH}} = c_{\mbox{\tiny GH}}(
    c_{\mbox{\tiny G}}^{\mbox{\tiny T}},
    c_{\mbox{\tiny H}}^{\mbox{\tiny T}}),
\hspace{0.2cm}
c_{\mbox{\tiny GHP}} = c_{\mbox{\tiny GHP}}(
     c_{\mbox{\tiny G}}^{\mbox{\tiny T}},
     c_{\mbox{\tiny H}}^{\mbox{\tiny T}}), \\
&& c_{\mbox{\tiny HP}} = c_{\mbox{\tiny HP}}(
     c_{\mbox{\tiny G}}^{\mbox{\tiny T}},
     c_{\mbox{\tiny H}}^{\mbox{\tiny T}}),
\hspace{0.2cm}
c_{\mbox{\tiny H}} = c_{\mbox{\tiny H}}(
    c_{\mbox{\tiny G}}^{\mbox{\tiny T}},
    c_{\mbox{\tiny H}}^{\mbox{\tiny T}}).
\end{eqnarray*}
We do not display these expressions here,
but they can be found in 
(\ref{eq:species_in_terms_of_totals})
of the Appendix. We note that these 
formulae can be used to calculate 
the equilibrium concentrations for 
the various species prior to release
since both
$c_{\mbox{\tiny G}}^{\mbox{\tiny T}}$
and
$c_{\mbox{\tiny H}}^{\mbox{\tiny T}}$ 
are known at $t=0$.  A numerical 
calculation is not required 
to obtain such quantities. In 
Figure \ref{fg:etag_etap}, we plot
curves for the equilibrium fraction 
of bound growth factor prior to release
using the formulae 
(\ref{eq:species_in_terms_of_totals}).
We also note that it is sufficient
to solve for 
$c_{\mbox{\tiny G}}^{\mbox{\tiny T}}$
and
$c_{\mbox{\tiny H}}^{\mbox{\tiny T}}$ 
for $t>0$ as the concentrations for 
$c_{\mbox{\tiny G}}$, $c_{\mbox{\tiny H}}$,
$c_{\mbox{\tiny GH}}$, $c_{\mbox{\tiny P}}$,
$c_{\mbox{\tiny HP}}$, $c_{\mbox{\tiny GHP}}$
then follow immediately from
(\ref{eq:species_in_terms_of_totals}). This 
implies that we can replace the problem 
containing five differential equations given
by 
(\ref{eq:nondimensional_governing_equations})
by the following problem containing just two
coupled partial differential equations:
\begin{eqnarray}
&& \frac{\partial 
   c_{\mbox{\tiny G}}^{\mbox{\tiny T}}}
   {\partial t} +
   \frac{\partial}{\partial x} \hspace{0.1cm}
   j_{\mbox{\tiny G}}^{\mbox{\tiny T}}
   (c_{\mbox{\tiny G}}^{\mbox{\tiny T}},
    c_{\mbox{\tiny H}}^{\mbox{\tiny T}},
    c_{\mbox{\tiny G}x}^{\mbox{\tiny T}},
    c_{\mbox{\tiny H}x}^{\mbox{\tiny T}})
  = 0, \hspace{0.2cm}  
   \frac{\partial 
   c_{\mbox{\tiny H}}^{\mbox{\tiny T}}}
   {\partial t} +
   \frac{\partial}{\partial x} \hspace{0.1cm}
   j_{\mbox{\tiny H}}^{\mbox{\tiny T}}
   (c_{\mbox{\tiny G}}^{\mbox{\tiny T}},
    c_{\mbox{\tiny H}}^{\mbox{\tiny T}},
    c_{\mbox{\tiny G}x}^{\mbox{\tiny T}},
    c_{\mbox{\tiny H}x}^{\mbox{\tiny T}})
  = 0, \nonumber \\
&& \frac{\partial
   c_{\mbox{\tiny G}}^{\mbox{\tiny T}}}
   {\partial x} = 0, \hspace{0.1cm}
   \frac{\partial
   c_{\mbox{\tiny H}}^{\mbox{\tiny T}}}
   {\partial x} = 0\hspace{0.1cm}\mbox{ on }
   x = 0, \label{eq:reduced_governing_system} \\ 
&& c_{\mbox{\tiny G}}^{\mbox{\tiny T}}
    = 0, \hspace{0.1cm}
   c_{\mbox{\tiny H}}^{\mbox{\tiny T}}
    = 0\hspace{0.1cm}\mbox{ on }x = 1, 
    \nonumber \\
&& c_{\mbox{\tiny G}}^{\mbox{\tiny T}}
    = 1, \hspace{0.1cm}
   c_{\mbox{\tiny H}}^{\mbox{\tiny T}}
    = 1
   \hspace{0.1cm}\mbox{ at }t = 0, \nonumber
\end{eqnarray}
where
\begin{eqnarray*}
&&  j_{\mbox{\tiny G}}^{\mbox{\tiny T}}
   (c_{\mbox{\tiny G}}^{\mbox{\tiny T}},
    c_{\mbox{\tiny H}}^{\mbox{\tiny T}},
    c_{\mbox{\tiny G}x}^{\mbox{\tiny T}},
    c_{\mbox{\tiny H}x}^{\mbox{\tiny T}})
    =
   - D_{\mbox{\tiny G}}^{*} 
   \frac{\partial}{\partial x} \hspace{0.1cm}
   c_{\mbox{\tiny G}}
   (c_{\mbox{\tiny G}}^{\mbox{\tiny T}},
    c_{\mbox{\tiny H}}^{\mbox{\tiny T}})
   - \frac{\partial}{\partial x} \hspace{0.1cm}
   c_{\mbox{\tiny GH}}
   (c_{\mbox{\tiny G}}^{\mbox{\tiny T}},
    c_{\mbox{\tiny H}}^{\mbox{\tiny T}}), \\
&&  \\
&& j_{\mbox{\tiny H}}^{\mbox{\tiny T}}
   (c_{\mbox{\tiny G}}^{\mbox{\tiny T}},
    c_{\mbox{\tiny H}}^{\mbox{\tiny T}},
    c_{\mbox{\tiny G}x}^{\mbox{\tiny T}},
    c_{\mbox{\tiny H}x}^{\mbox{\tiny T}})
    =
   - D_{\mbox{\tiny H}}^{*} 
   \frac{\partial}{\partial x} \hspace{0.1cm}
   c_{\mbox{\tiny H}}
   (c_{\mbox{\tiny G}}^{\mbox{\tiny T}},
    c_{\mbox{\tiny H}}^{\mbox{\tiny T}})
   - \frac{1}{\eta_{\mbox{\tiny H/G}}}
   \frac{\partial}{\partial x} \hspace{0.1cm}
   c_{\mbox{\tiny GH}}
   (c_{\mbox{\tiny G}}^{\mbox{\tiny T}},
    c_{\mbox{\tiny H}}^{\mbox{\tiny T}}),     
\end{eqnarray*}
and where the expressions for 
$c_{\mbox{\tiny G}}
   (c_{\mbox{\tiny G}}^{\mbox{\tiny T}},
    c_{\mbox{\tiny H}}^{\mbox{\tiny T}})$
and
$c_{\mbox{\tiny GH}}
   (c_{\mbox{\tiny G}}^{\mbox{\tiny T}},
    c_{\mbox{\tiny H}}^{\mbox{\tiny T}})$    
are given in 
(\ref{eq:species_in_terms_of_totals}). We note
that (\ref{eq:reduced_governing_system}) is in 
a standard form that can be readily given to a 
mathematical package such as MAPLE to solve.   

\section{Analysis and results}

\subsection{Optimal conditions for 
slow passive release: strongly retained
heparin and growth factor}
 
In a medical device such as a nerve 
guide tube, it is frequently required 
to maintain growth factor in the 
device until such time as it is actively 
released by invading cells. In such cases,
the device should be designed so as to 
minimise passive release of growth 
factor via diffusion. There are five 
dimensionless parameters that can in 
principle be independently varied in 
experiments to tune the system for 
a {\em given} growth factor, and these 
are:
\[
\eta_{\mbox{\tiny H/G}} = 
 \frac{c_{\mbox{\tiny H}}^{\mbox{\tiny 0}}}
 {c_{\mbox{\tiny G}}^{\mbox{\tiny 0}}}, 
 \hspace{0.2cm}
 \eta_{\mbox{\tiny H/P}} = 
 \frac{c_{\mbox{\tiny H}}^{\mbox{\tiny 0}}}
 {c_{\mbox{\tiny P}}^{\mbox{\tiny 0}}},
 \hspace{0.2cm}
 \theta = \frac{k_{r}}
{{\mbox{\tiny K}}_{\mbox{\tiny R}}},
 \hspace{0.2cm} 
 K_{b{\mbox{\tiny H}}} = 
\frac{c_{\mbox{\tiny H}}^{\mbox{\tiny 0}}}
{K_{\mbox{\tiny G-H}}^{\mbox{\tiny D}}}, 
\hspace{0.2cm}
K_{b{\mbox{\tiny P}}} = 
\frac{c_{\mbox{\tiny P}}^{\mbox{\tiny 0}}}
{K_{\mbox{\tiny H-P}}^{\mbox{\tiny D}}}.
%
\]
We emphasise that the parameters
$K_{\mbox{\tiny H-P}}^{\mbox{\tiny D}}$
and $\kappa_{\mbox{\tiny R}}$ are in 
principle tunable since peptides with 
desired properties can be designed 
(\cite{Wood-2007}). However, if the peptide 
is also fixed, only three dimensionless 
parameters can be independently varied 
in the experiments, one possible choice 
for these being 
$\eta_{\mbox{\tiny H/G}},
\eta_{\mbox{\tiny H/P}}$ and
$K_{b{\mbox{\tiny H}}}$.
The parameters 
$D_{\mbox{\tiny G}}^{*}$ and 
$D_{\mbox{\tiny H}}^{*}$ cannot be
changed in experiments as they are 
fixed for a given growth factor. The 
parameters $\delta_{\mbox{\tiny G}}$ and
$\delta_{\mbox{\tiny H}}$ are neglected
here since they are typically tiny in 
systems of interest.  

In the literature to date, the emphasis
has been on experimentally varying the 
parameters $\eta_{\mbox{\tiny H/G}}$ and
$\eta_{\mbox{\tiny H/P}}$ to determine 
optimal conditions for slow passive
release; see \cite{S-Elbert-2000a, 
Taylor-2004, Willerth-2008} and 
\cite{Wood-2007, Wood-2008}. In 
particular, experiments have been 
carried out for very large values 
of the ratio $\eta_{\mbox{\tiny H/G}}$, 
and quite small values for the 
ratio $\eta_{\mbox{\tiny H/P}}$.
However, we now show that if one 
wishes to ensure slow passive 
release, then the key parameters 
to monitor are
$K_{b{\mbox{\tiny H}}}$ and
$K_{b{\mbox{\tiny P}}}$, rather 
than
$\eta_{\mbox{\tiny H/G}}$ and
$\eta_{\mbox{\tiny H/P}}$. More 
precisely, we shall show that 
slow release of at least a 
proportion of the growth factor 
is assured provided 
$K_{b{\mbox{\tiny H}}},
K_{b{\mbox{\tiny P}}} \gg 1$ 
(with the other parameters being 
$O(1)$, although there are other
possibilities), or, equivalently:
\begin{equation} 
c_{\mbox{\tiny H}}^{\mbox{\tiny 0}} \gg
K_{\mbox{\tiny G-H}}^{\mbox{\tiny D}}
\hspace{0.2cm}\mbox{ and }\hspace{0.2cm}
c_{\mbox{\tiny P}}^{\mbox{\tiny 0}} \gg
K_{\mbox{\tiny H-P}}^{\mbox{\tiny D}}.
\label{eq:criteria}
\end{equation}
We recall that 
$K_{b{\mbox{\tiny H}}},
K_{b{\mbox{\tiny P}}} \gg 1$ 
corresponds to strong retention 
of both growth factor by the 
heparin, and of heparin by the 
peptide. Hence, if practicable, 
for slow release of growth 
factor, the matrix should usually 
be prepared with the initial 
concentration of heparin being 
much larger than the dissociation 
constant of growth factor from 
heparin, and the concentration 
of peptide covalently cross-linked 
to the fibrin matrix peptide being 
much larger than the dissociation 
constant of heparin from peptide.  
We now justify this conclusion 
using an asymptotic argument and 
by providing numerical evidence. 
In particular, we shall demonstrate 
numerically that growth factor 
release can be relatively fast if 
the conditions (\ref{eq:criteria}) 
are {\em not} met even with 
$\eta_{\mbox{\tiny H/G}} \gg 1$
and
$\eta_{\mbox{\tiny H/P}} \ll 1$.

\subsubsection{Asymptotics:
$K_{b{\mbox{\tiny H}}},
K_{b{\mbox{\tiny P}}} \gg 1$} 

\begin{figure}[!t]
\centerline{\includegraphics [width=8cm,height=6.5cm]{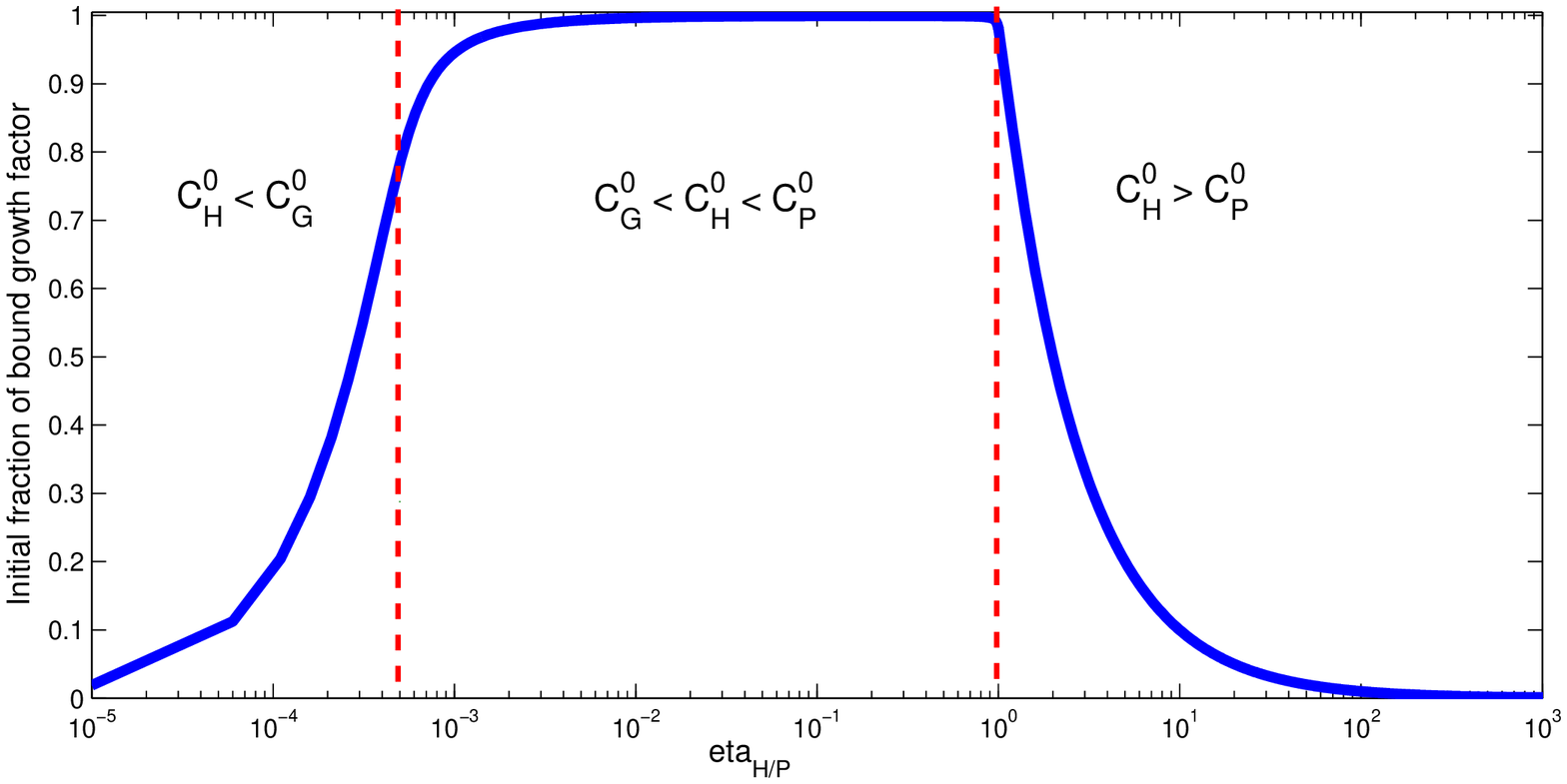}   
 \hskip -0.8cm
    \includegraphics [width=8cm,height=6.5cm]{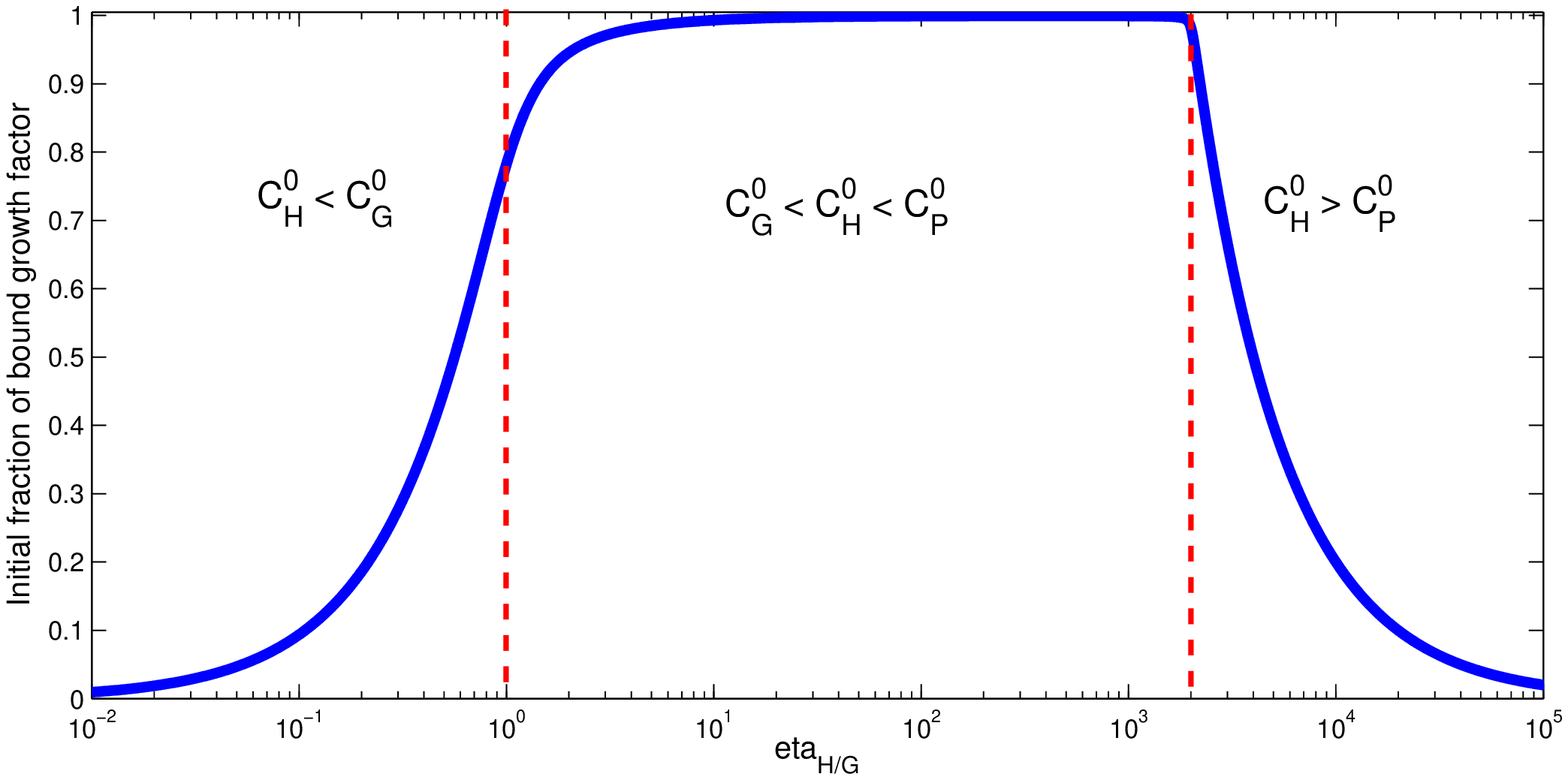}}  
      \vskip -0.2cm
      \caption{Theoretical curves for the 
      equilibrium bound fraction of growth 
      factor prior to release. The curves are 
      calculated using the formulae 
      (\ref{eq:species_in_terms_of_totals}).     
      In the figure on the left, 
      $\eta_{\mbox{\tiny H/P}}$ is varied 
      and the remaining non-dimensional 
      parameters are fixed. In the figure
      on the right, $\eta_{\mbox{\tiny H/G}}$ 
      is varied with the other parameters 
      fixed. The parameter values used are 
      $D_{\mbox{\tiny G}}^*=6$, 
      $D_{\mbox{\tiny H}}^*=3$, $\eta_{\mbox{\tiny H/G}}/\eta_{\mbox{\tiny H/P}}=2000$, $\theta = 9 \times 10^{-3}$, 
      $K_{b{\mbox{\tiny P}}}=3000$ and 
      $K_{\mbox{\tiny G-H}}^{\mbox{\tiny D}}=
      7.5 \times 10^{-9}$M (\cite{Nugent-1992}).} 
    \label{fg:etag_etap}     
\end{figure}

We write
$K_{b{\mbox{\tiny H}}} = 1/\varepsilon$,
$K_{b{\mbox{\tiny P}}} = \mu /\varepsilon$
and consider the limit 
$\varepsilon\rightarrow 0$ in
(\ref{eq:species_in_terms_of_totals})
for fixed $O(1)$ values of
$c_{\mbox{\tiny G}}^{\mbox{\tiny T}}$
and 
$c_{\mbox{\tiny H}}^{\mbox{\tiny T}}$, 
and with $\mu$ and all the remaining 
dimensionless parameters in
(\ref{eq:species_in_terms_of_totals})
being $O(1)$. The fraction of bound 
drug in the matrix, which we denote 
by $f_{\mbox{\tiny B}}$, is given by:
\[ 
f_{\mbox{\tiny B}}=
\frac{c_{\mbox{\tiny GHP}}}
{c_{\mbox{\tiny G}} 
 +  c_{\mbox{\tiny GH}} 
 +  c_{\mbox{\tiny GHP}}}=
 \frac{c_{\mbox{\tiny GHP}}}
 {c_{\mbox{\tiny G}}
  ^{\mbox{\tiny T}}}; 
\]
some solutions for this quantity 
at $t=0$ are displayed in Figure 
\ref{fg:etag_etap}. For clarity, 
we revert to dimensional quantities 
in this Section. In the limit 
$\varepsilon\rightarrow 0$, we find 
that: 
\begin{equation}
f_{\mbox{\tiny B}} \sim 
\left\{ 
       \begin{array}{ll}
\frac{ 
 c_{\mbox{\tiny H}}^{\mbox{\tiny T}}}
 {c_{\mbox{\tiny G}}^{\mbox{\tiny T}}} 
 & \mbox{ if }  
 c_{\mbox{\tiny H}}^{\mbox{\tiny T}} 
 < c_{\mbox{\tiny G}}^{\mbox{\tiny T}} 
 \mbox{ and } 
 c_{\mbox{\tiny H}}^{\mbox{\tiny T}} < 
  c_{\mbox{\tiny P}}^{\mbox{\tiny 0}}, \\ \\
 
 \frac{c_{\mbox{\tiny P}}^{\mbox{\tiny 0}}}
  {c_{\mbox{\tiny G}}^{\mbox{\tiny T}}}
 
  & \mbox{ if }  
 c_{\mbox{\tiny H}}^{\mbox{\tiny T}} 
 < c_{\mbox{\tiny G}}^{\mbox{\tiny T}} 
 \mbox{ and } 
 c_{\mbox{\tiny H}}^{\mbox{\tiny T}} > 
 c_{\mbox{\tiny P}}^{\mbox{\tiny 0}}, \\ \\
  \frac{c_{\mbox{\tiny P}}^{\mbox{\tiny 0}}}
 {c_{\mbox{\tiny H}}^{\mbox{\tiny T}}}

  & \mbox{ if }  
 c_{\mbox{\tiny H}}^{\mbox{\tiny T}} 
 > c_{\mbox{\tiny G}}^{\mbox{\tiny T}} 
 \mbox{ and } 
 c_{\mbox{\tiny H}}^{\mbox{\tiny T}} > 
 c_{\mbox{\tiny P}}^{\mbox{\tiny 0}},\\  \\
 1
 
 & \mbox{ if }  
 c_{\mbox{\tiny H}}^{\mbox{\tiny T}} 
 > c_{\mbox{\tiny G}}^{\mbox{\tiny T}} 
 \mbox{ and } 
 c_{\mbox{\tiny H}}^{\mbox{\tiny T}} < 
 c_{\mbox{\tiny P}}^{\mbox{\tiny 0}}.
\end{array} \right.
\label{eq:fBregimes}
\end{equation}
There are also three narrow transition
regions at the interfaces of the
four regimes listed above, but we omit
this detail since it does not 
contribute to the subsequent discussion.
The results of (\ref{eq:fBregimes}) are
readily interpreted. Take, for example,
the case $f_{\mbox{\tiny B}}\sim 1$
for
$c_{\mbox{\tiny H}}^{\mbox{\tiny T}} 
 > c_{\mbox{\tiny G}}^{\mbox{\tiny T}}$ 
and  
$c_{\mbox{\tiny H}}^{\mbox{\tiny T}} < 
 c_{\mbox{\tiny P}}^{\mbox{\tiny 0}}$.
Since in the current limit both the growth 
factor and heparin are strongly 
retained, then provided there is 
enough heparin to accomodate the 
growth factor 
($c_{\mbox{\tiny H}}^{\mbox{\tiny T}} 
 > c_{\mbox{\tiny G}}^{\mbox{\tiny T}}$)
and enough peptide to accomodate the 
heparin 
($c_{\mbox{\tiny H}}^{\mbox{\tiny T}} 
< c_{\mbox{\tiny P}}^{\mbox{\tiny 0}}$),
all of the growth factor in the 
matrix will be bound to leading order
($f_{\mbox{\tiny B}}\sim 1$). This is
the desired regime for slow passive
release, as we now confirm. The other 
three cases are similarly interpreted.

To gain insight into the time it would 
take for the growth factor to passively
release from the matrix, we now consider
the total flux of growth factor, 
$j_{\mbox{\tiny G}}^{\mbox{\tiny T}}$.
We find as $\varepsilon \rightarrow 0$ 
that:
\begin{equation}
j_{\mbox{\tiny G}}^{\mbox{\tiny T}} \sim 
\left\{ 
       \begin{array}{ll}
 - D_{\mbox{\tiny G}} \left( 
  \frac{\partial c_{\mbox{\tiny G}}^{\mbox{\tiny T}}}
  {\partial x} -
  \frac{c_{\mbox{\tiny G}}^{\mbox{\tiny 0}}}
  {c_{\mbox{\tiny H}}^{\mbox{\tiny 0}}   } 
  \frac{\partial c_{\mbox{\tiny H}}^{\mbox{\tiny T}}}
 {\partial x} \right) 
 & \mbox{ if }  
 c_{\mbox{\tiny H}}^{\mbox{\tiny T}} 
 < c_{\mbox{\tiny G}}^{\mbox{\tiny T}} 
 \mbox{ and } 
 c_{\mbox{\tiny H}}^{\mbox{\tiny T}} < 
 c_{\mbox{\tiny P}}^{\mbox{\tiny 0}}, \\ \\
 -D_{\mbox{\tiny G}} 
 \frac{\partial c_{\mbox{\tiny G}}^{\mbox{\tiny T}}}
 {\partial x} - 
 (D_{\mbox{\tiny GH}} -D_{\mbox{\tiny G}})
 \frac{\partial c_{\mbox{\tiny H}}^{\mbox{\tiny T}}}
 {\partial x}  
 & \mbox{ if }  
 c_{\mbox{\tiny H}}^{\mbox{\tiny T}} 
 < c_{\mbox{\tiny G}}^{\mbox{\tiny T}} 
 \mbox{ and } 
 c_{\mbox{\tiny H}}^{\mbox{\tiny T}} > 
 c_{\mbox{\tiny P}}^{\mbox{\tiny 0}}, \\ \\
 -D_{\mbox{\tiny GH}} \left(
 \left\{ \frac{c_{\mbox{\tiny H}}^{\mbox{\tiny T}} 
 - c_{\mbox{\tiny P}}^{\mbox{\tiny 0}}}
 { c_{\mbox{\tiny H}}^{\mbox{\tiny T}}} \right\} 
 \frac{\partial c_{\mbox{\tiny G}}^{\mbox{\tiny T}}}
 {\partial x} + 
 \frac{ c_{\mbox{\tiny P}}^{\mbox{\tiny 0}}
 c_{\mbox{\tiny G}}^{\mbox{\tiny T}}}
 {{c_{\mbox{\tiny H}}^{\mbox{\tiny T}}}^2} 
 \frac{\partial c_{\mbox{\tiny H}}^{\mbox{\tiny T}}}
 {\partial x} \right)
 & \mbox{ if }  
 c_{\mbox{\tiny H}}^{\mbox{\tiny T}} 
 > c_{\mbox{\tiny G}}^{\mbox{\tiny T}} 
 \mbox{ and } 
 c_{\mbox{\tiny H}}^{\mbox{\tiny T}} > 
 c_{\mbox{\tiny P}}^{\mbox{\tiny 0}},\\  \\
 - \varepsilon \left( 
 A(c_{\mbox{\tiny G}}^{\mbox{\tiny T}},
 c_{\mbox{\tiny H}}^{\mbox{\tiny T}})
 \frac{\partial c_{\mbox{\tiny G}}
 ^{\mbox{\tiny T}}}
 {\partial x} 
 + B(c_{\mbox{\tiny G}}^{\mbox{\tiny T}},
 c_{\mbox{\tiny H}}^{\mbox{\tiny T}})
 \frac{\partial c_{\mbox{\tiny H}}^{\mbox{\tiny T}}}
 {\partial x} \right)
 & \mbox{ if }  
 c_{\mbox{\tiny H}}^{\mbox{\tiny T}} 
 > c_{\mbox{\tiny G}}^{\mbox{\tiny T}} 
 \mbox{ and } 
 c_{\mbox{\tiny H}}^{\mbox{\tiny T}} < 
 c_{\mbox{\tiny P}}^{\mbox{\tiny 0}},
\end{array} \right.  \label{eq:jGTregimes}
\end{equation}
where: 
\[
A(c_{\mbox{\tiny G}}^{\mbox{\tiny T}},
 c_{\mbox{\tiny H}}^{\mbox{\tiny T}}) 
 = \frac{c_{\mbox{\tiny H}}^{\mbox{\tiny 0}}
    c_{\mbox{\tiny H}}^{\mbox{\tiny T}}}
  {(c_{\mbox{\tiny G}}^{\mbox{\tiny T}} -
  c_{\mbox{\tiny H}}^{\mbox{\tiny T}})^{2}} 
  D_{\mbox{\tiny G}}
 -\frac{ \mu c_{\mbox{\tiny P}}^{\mbox{\tiny 0}}}
 {c_{\mbox{\tiny H}}^{\mbox{\tiny T}}  
 -c_{\mbox{\tiny P}}^{\mbox{\tiny 0}}} 
 D_{\mbox{\tiny GH}},
 \hspace{0.2cm}
B(c_{\mbox{\tiny G}}^{\mbox{\tiny T}},
 c_{\mbox{\tiny H}}^{\mbox{\tiny T}})
 = \frac{\mu c_{\mbox{\tiny P}}^{\mbox{\tiny 0}}
 c_{\mbox{\tiny G}}^{\mbox{\tiny T}}}
 { (c_{\mbox{\tiny H}}^{\mbox{\tiny T}}  
 -c_{\mbox{\tiny P}}^{\mbox{\tiny 0}})^{2}} 
 D_{\mbox{\tiny GH}}
 -\frac{ c_{\mbox{\tiny H}}^{\mbox{\tiny 0}}
 c_{\mbox{\tiny G}}^{\mbox{\tiny T}} }
 {(c_{\mbox{\tiny G}}^{\mbox{\tiny T}} -
  c_{\mbox{\tiny H}}^{\mbox{\tiny T}})^{2}} 
 D_{\mbox{\tiny G}}, 
\]
and assuming the dimensionless form for the 
derivatives arising are $O(1)$. The point 
to note here is that the flux of growth 
factor for the fourth regime,
$c_{\mbox{\tiny H}}^{\mbox{\tiny T}} 
 > c_{\mbox{\tiny G}}^{\mbox{\tiny T}}$ 
and  
$c_{\mbox{\tiny H}}^{\mbox{\tiny T}} < 
 c_{\mbox{\tiny P}}^{\mbox{\tiny 0}}$,
is $O(\varepsilon)$ smaller than that for
the other three. 

In view of 
(\ref{eq:fBregimes}) and 
(\ref{eq:jGTregimes}), it is now 
clear that the optimal regime for 
slow passive release is
$c_{\mbox{\tiny H}}^{\mbox{\tiny T}} 
 > c_{\mbox{\tiny G}}^{\mbox{\tiny T}}$ 
and  
$c_{\mbox{\tiny H}}^{\mbox{\tiny T}} < 
 c_{\mbox{\tiny P}}^{\mbox{\tiny 0}}$,
which indicates that the polymerization
mixture for the matrix should have
$c_{\mbox{\tiny G}}^{\mbox{\tiny 0}} < 
 c_{\mbox{\tiny H}}^{\mbox{\tiny 0}} <
 c_{\mbox{\tiny P}}^{\mbox{\tiny 0}}$.
Hence, we expand our original recommendations 
for matrix preparation, (\ref{eq:criteria}), to 
the following:
\begin{equation} 
c_{\mbox{\tiny H}}^{\mbox{\tiny 0}} \gg
K_{\mbox{\tiny G-H}}^{\mbox{\tiny D}},
\hspace{0.2cm}
c_{\mbox{\tiny P}}^{\mbox{\tiny 0}} \gg
K_{\mbox{\tiny H-P}}^{\mbox{\tiny D}},
\hspace{0.2cm}
c_{\mbox{\tiny G}}^{\mbox{\tiny 0}} < 
 c_{\mbox{\tiny H}}^{\mbox{\tiny 0}} <
 c_{\mbox{\tiny P}}^{\mbox{\tiny 0}}.
\label{eq:full_criteria}
\end{equation}
%


We have a `large' growth factor flux
for the first three cases in 
(\ref{eq:jGTregimes}) because for each 
of these regimes, there is a substantial
free component that can diffuse. For 
example, for the first case,  
$c_{\mbox{\tiny H}}^{\mbox{\tiny T}} 
 < c_{\mbox{\tiny G}}^{\mbox{\tiny T}}$
and 
$c_{\mbox{\tiny H}}^{\mbox{\tiny T}} < 
c_{\mbox{\tiny P}}^{\mbox{\tiny 0}}$,
we have
$c_{\mbox{\tiny G}} \sim 
c_{\mbox{\tiny G}}^{\mbox{\tiny T}} 
- c_{\mbox{\tiny H}}^{\mbox{\tiny T}}$.
For the boundary and initial conditions
of (\ref{eq:reduced_governing_system}),
this free drug will clear the system to
leading order on the time scale 
$t=O(L^{2}/D_{\mbox{\tiny GH}})$. 
However, once the free component has
cleared, the remaining bound component
in the bulk will be governed by the 
fourth regime of (\ref{eq:jGTregimes}), 
and this will clear the system on the 
long time scale
$t=O(L^{2}/(\varepsilon 
D_{\mbox{\tiny GH}}))
\gg O(L^{2}/ 
D_{\mbox{\tiny GH}})$. 
Similar remarks apply to the second and 
third regimes in (\ref{eq:jGTregimes}). 
Hence, if the matrix is prepared with
$K_{b{\mbox{\tiny H}}}, 
K_{b{\mbox{\tiny P}}}\gg 1$, and if our 
governing model is appropriate, then 
we are assured that at least a fraction
of the growth factor will release on the
slow time scale
$t=O(L^{2}/(\varepsilon 
D_{\mbox{\tiny GH}}))$. Furthermore, 
we predict that almost all of the
growth factor will release slowly if 
the matrix is prepared with 
$c_{\mbox{\tiny H}}^{\mbox{\tiny 0}}/
c_{\mbox{\tiny G}}^{\mbox{\tiny 0}}>1$
and
$c_{\mbox{\tiny P}}^{\mbox{\tiny 0}}/
c_{\mbox{\tiny H}}^{\mbox{\tiny 0}}>1$;
notice that it is not required that 
these fractions be large; see 
Figure \ref{fg:etag_etap}. However,
we should caution that if there is 
a substantial component of free 
peptide (which is not included in 
the model described here), then 
there can still be a significant 
amount of free growth factor that 
can release on a fast diffusion 
time scale.  

The two stage release behaviour just 
described has been observed in 
experiments 
(see Section \ref{sec:comparison}), 
where one sometimes sees a proportion 
of the growth factor releasing quickly 
over a period of some days (which could 
correspond to free growth factor releasing 
on a diffusion time scale) followed by 
much slower release of the remaining 
fraction (which could correspond to a 
strongly retained bound component 
releasing on a longer time scale such 
as that described above). 
 
\vspace{0.3cm}

\noindent {\em Incorporating free 
peptide in the analysis}

\vspace{0.3cm}

We now consider the case where a 
substantial fraction of the peptide 
remains free and competes with the
covalently bound peptide for free
heparin. We assume that heparin bound
to free peptide has the same binding
behaviour for growth factor as free 
heparin. The essence of our results 
above carry over, as we now explain. 
We suppose that the conditions 
(\ref{eq:full_criteria})
hold, and that the ratio of peptide 
covalently attached to the fibrin, 
$r$, has been quantified. Then the 
concentration of cross-linked peptide 
in the system is 
$r c_{\mbox{\tiny P}}^{\mbox{\tiny 0}}$,
and since 
$c_{\mbox{\tiny P}}^{\mbox{\tiny 0}}>
c_{\mbox{\tiny H}}^{\mbox{\tiny 0}}$,
the initial concentration of heparin 
that is bound to cross-linked peptide 
is, to leading order, 
$r c_{\mbox{\tiny H}}^{\mbox{\tiny 0}}$.
Since in turn  
$c_{\mbox{\tiny H}}^{\mbox{\tiny 0}}>
c_{\mbox{\tiny G}}^{\mbox{\tiny 0}}$,
the initial concentration of growth 
factor trapped by the delivery system 
is, to leading order, 
$r c_{\mbox{\tiny G}}^{\mbox{\tiny 0}}$.
It follows that a fraction $(1-r)$ 
approximately of the growth factor 
will diffuse out of the system over 
a diffusion time scale. Hence, the 
final recommendation, which we add 
to (\ref{eq:full_criteria}), is that 
$c_{\mbox{\tiny G}}^{\mbox{\tiny 0}}$ 
should be chosen so that
$r c_{\mbox{\tiny G}}^{\mbox{\tiny 0}}$ 
is sufficiently large for the therapy 
to be effective.
 
\subsubsection{Numerical Solutions}

Two different procedures were used to 
numerically integrate the initial 
boundary value problem
(\ref{eq:reduced_governing_system}). In 
one method, simple explicit time-stepping 
was used to update the values of 
$c_{\mbox{\tiny G}}^{\mbox{\tiny T}}$,
$c_{\mbox{\tiny H}}^{\mbox{\tiny T}}$,
with the other quantities being then 
updated using
(\ref{eq:species_in_terms_of_totals}). 
Centred difference approximations 
were used for
$c_{\mbox{\tiny G}xx}$, 
$c_{\mbox{\tiny H}xx}$,
$c_{\mbox{\tiny GH}xx}$, and the 
no-flux conditions on $x=0$ were 
handled by introducing a fictitious 
line in the usual way. In the other 
method, the system was numerically 
integrated using the MAPLE command 
\ttfamily{pdsolve/numeric}, \normalfont
which is based on a centred implict 
finite difference scheme. Good 
agreement was obtained between the 
two schemes and with known analytical 
results.


\begin{figure}[!t]
\centerline{\includegraphics [width=13.5cm,height=8.5cm]{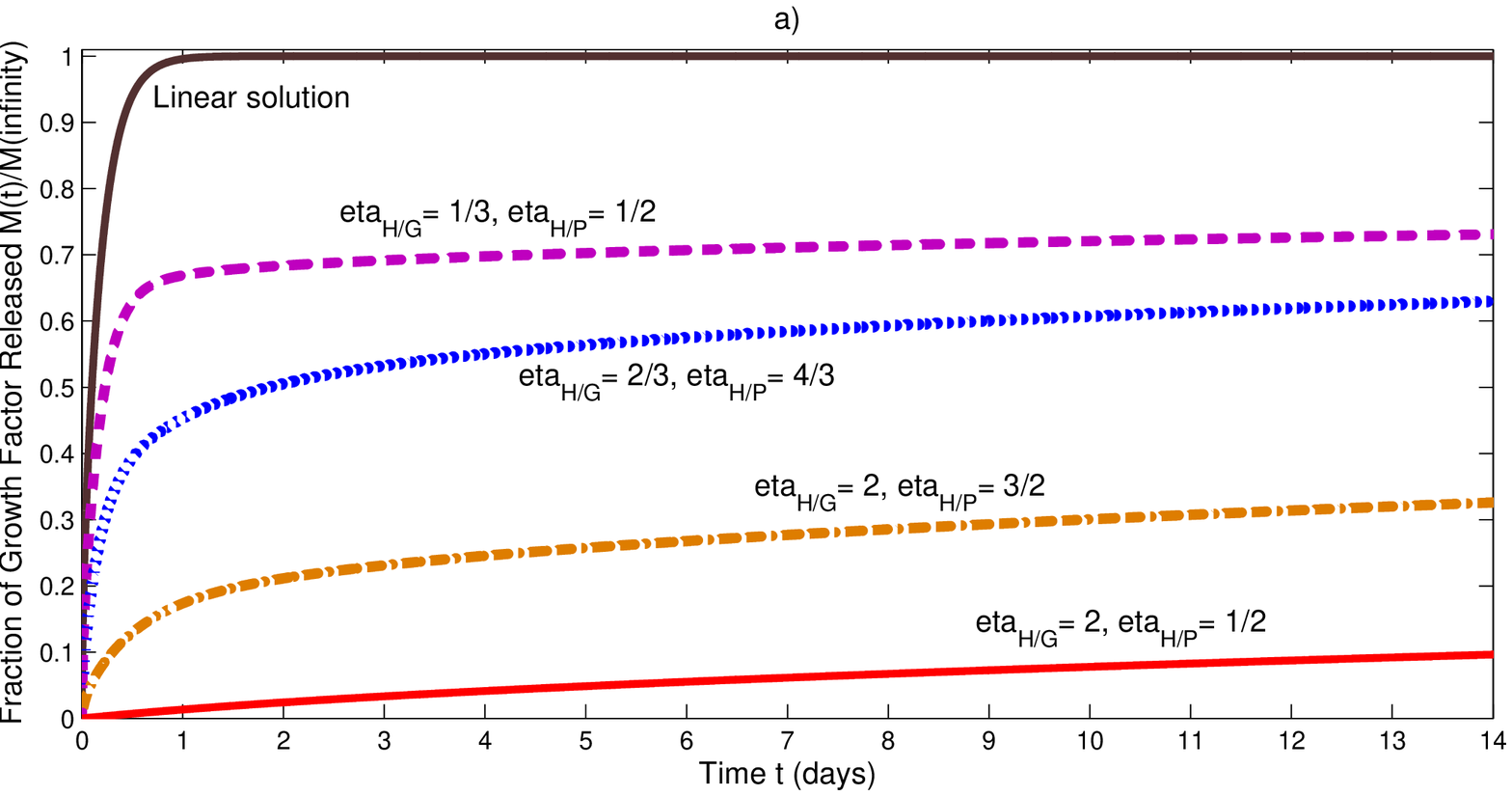}}   
 \centerline{    \includegraphics [width=14cm,height=8.5cm]{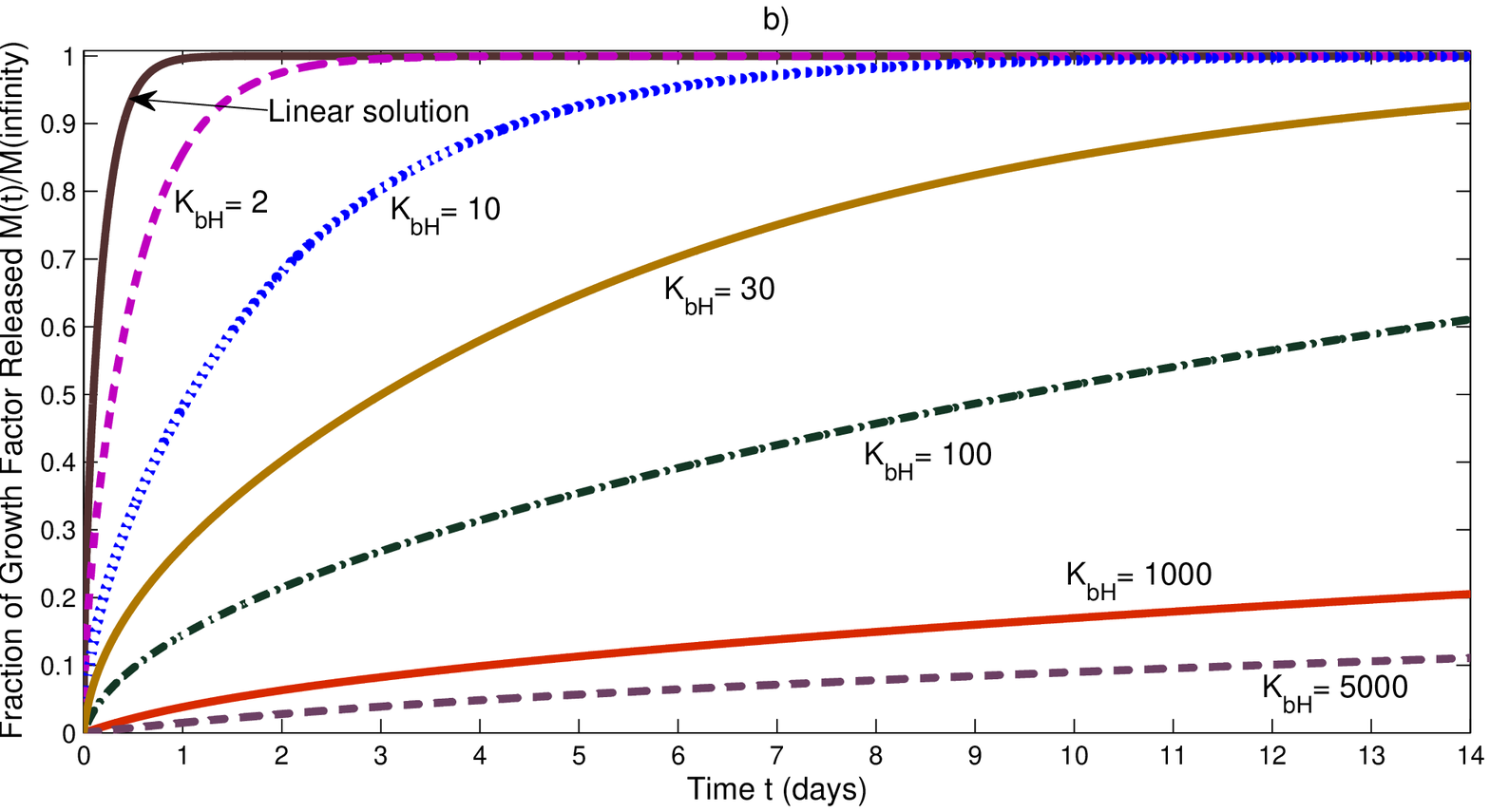}}  
      \vskip -0.2cm
      \caption{Numerical solutions 
      of the initial boundary value 
      problem
      (\ref{eq:reduced_governing_system}). 
      We display the predicted 
      fraction of total growth  
      factor released over a two 
      week period. We use the 
      parameter values
      $D_{\mbox{\tiny G}}^*=6$, 
      $D_{\mbox{\tiny H}}^*=3$, $\theta = 9 \times 10^{-3}$, 
      $K_{b{\mbox{\tiny P}}}=3000$
      throughout. In (a), 
      $K_{b{\mbox{\tiny H}}}=8000$
      with various
      $\eta_{\mbox{\tiny H/G}}$,
      $\eta_{\mbox{\tiny H/P}}$ 
      values indicated on the curves. 
      In (b),  
      $\eta_{\mbox{\tiny H/G}}=1000$, 
      $\eta_{\mbox{\tiny H/P}}= 1/40$
      with various
      $K_{b{\mbox{\tiny H}}}$ 
      values indicated on the curves.} 
 \label{fg:vary_parameters}     
\end{figure}


\begin{figure}[!t]
\vskip -0.2cm
\centerline{\includegraphics [width=12cm,height=6.8cm]{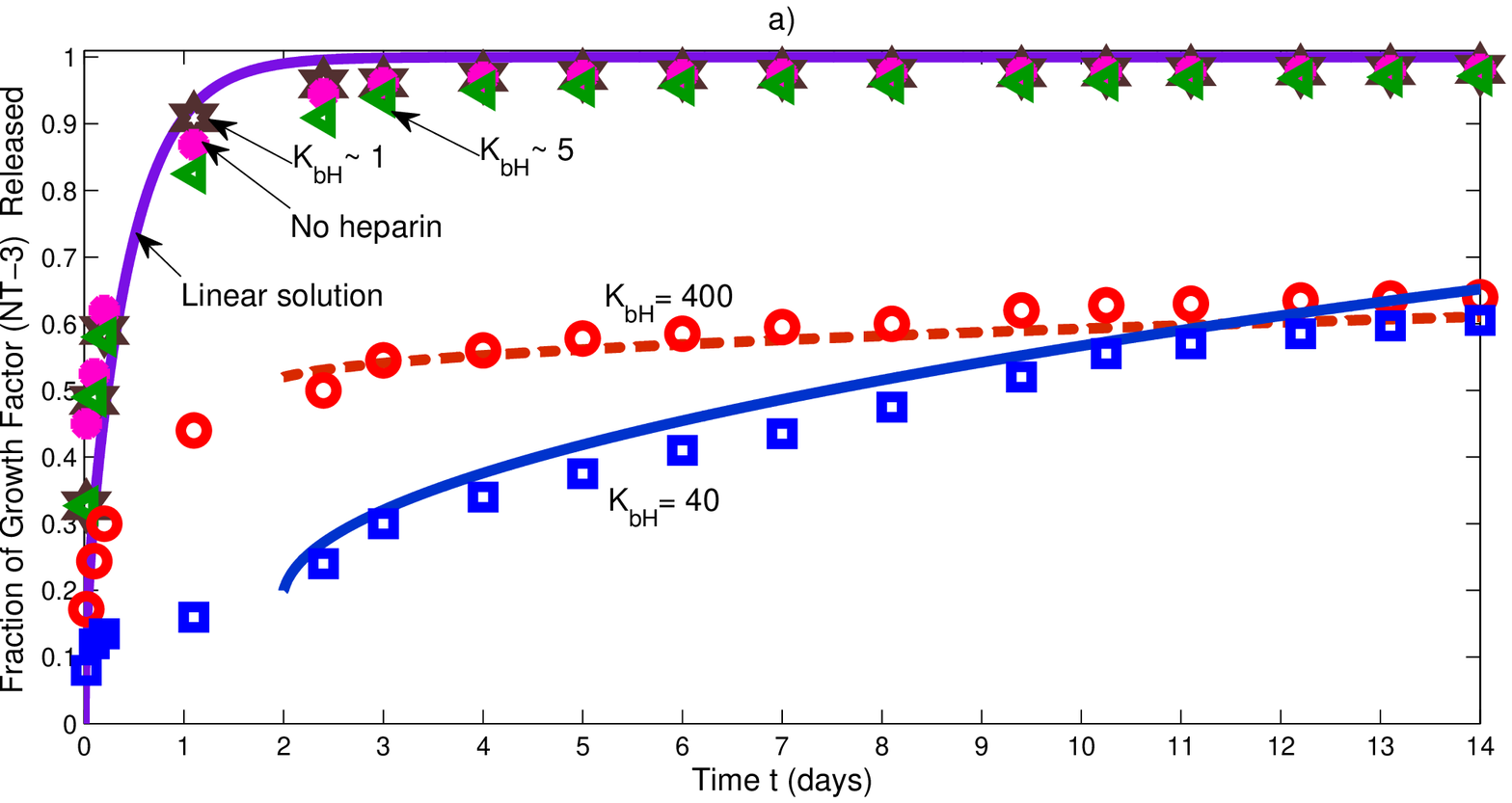}}
\vskip -0.2cm
\centerline{  \includegraphics [width=11.5cm,height=6.8cm]{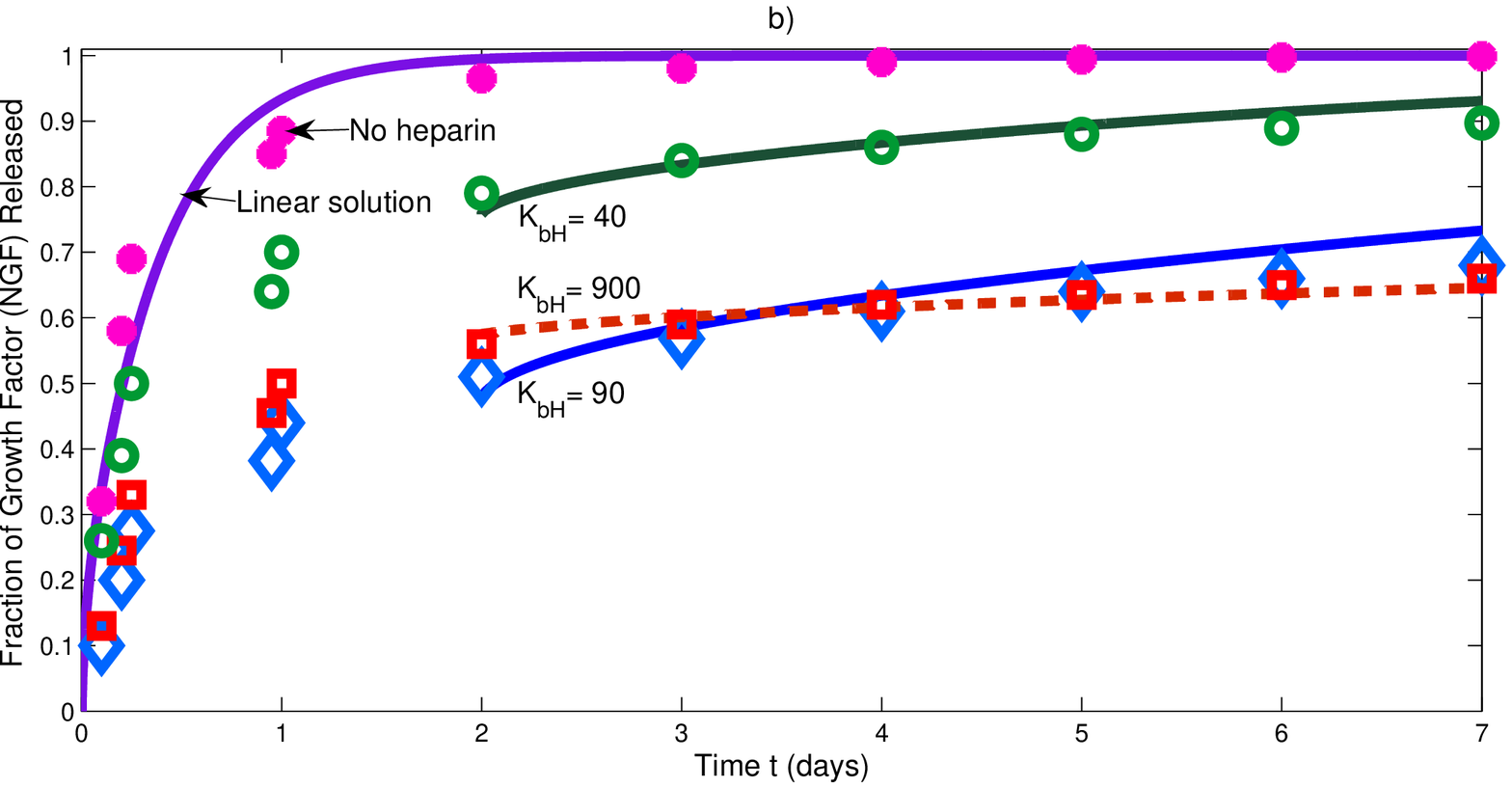}}
\vskip -0.2cm  
 \centerline{ \includegraphics [width=11.5cm,height=6.8cm]{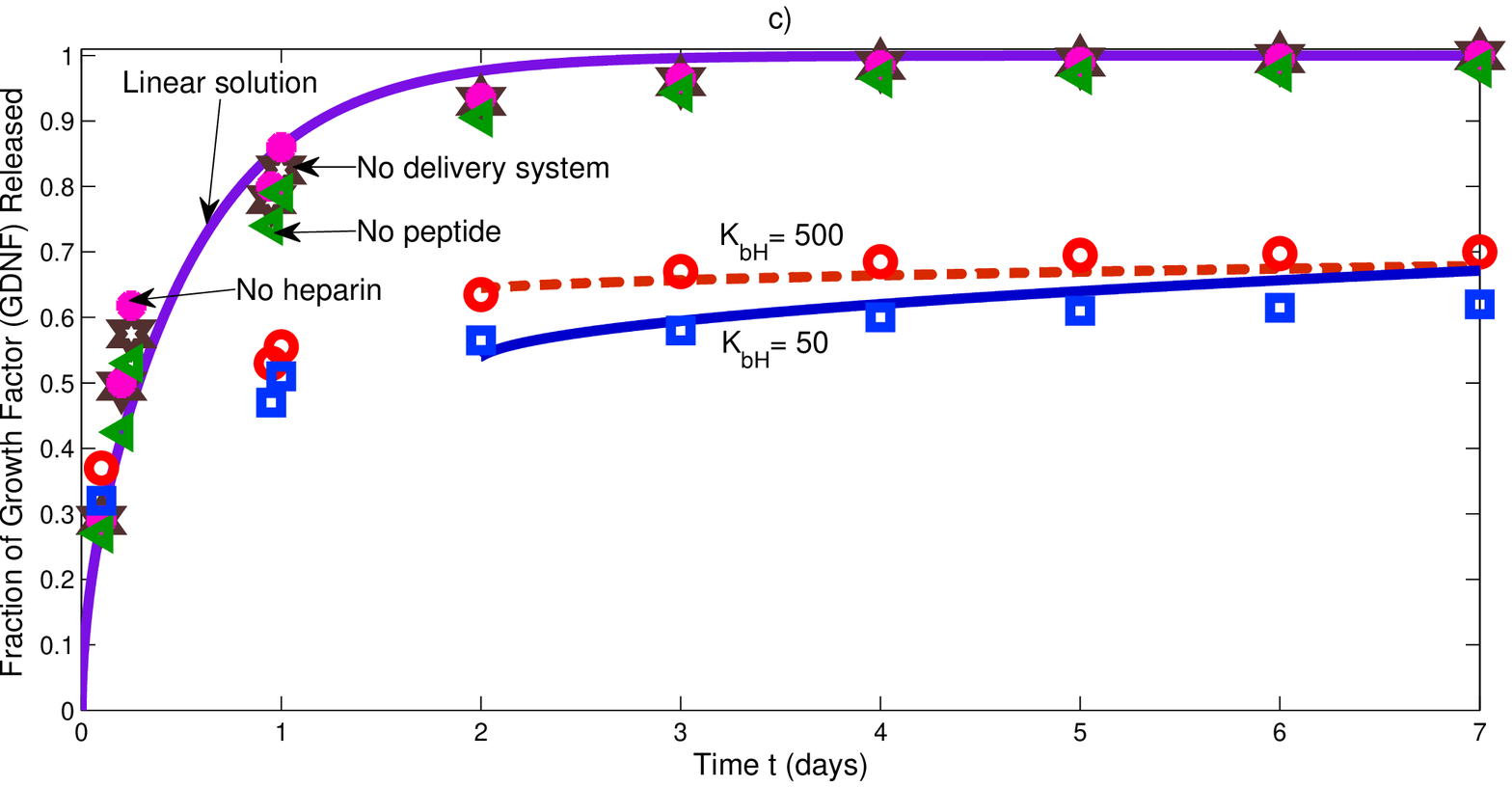}}    
      \vskip -0.2cm
      \caption{Comparison of experimental and 
         theoretical release profiles. 
         The curves are theoretical
         and the symbols are experimental. 
         In the figure, 
         (a) has data for NT-3 taken 
         from \cite{Taylor-2004}, 
         (b) has data for NGF taken from 
         \cite{Wood-2007}, and 
         (c) has data for GDNF taken from 
         \cite{Wood-2008}. The values of  
         $K_{b{\mbox{\tiny H}}}$ used to
         generate the theoretical curves
         are given on them and the remaining 
         parameters used can be found in 
         Table \ref{table:fitting} of the 
         Appendix. In (a), two approximate 
         $K_{b{\mbox{\tiny H}}}$
         values for experimental data are 
         also given.} 
 \label{fg:fitting}     
\end{figure}


In Figure \ref{fg:vary_parameters}, 
we display numerical profiles for 
the fraction of total 
growth factor that has released 
from the system as a function of 
time over a period of a fortnight. 
Four of the five release 
profiles displayed in 
Figure \ref{fg:vary_parameters} (a) 
correspond to the case 
$K_{b{\mbox{\tiny H}}}, 
K_{b{\mbox{\tiny P}}}\gg 1$, the 
regime we recommend for matrix 
preparation. The other parameter
values are all $O(1)$, and can 
be found either on the figure or in 
its caption. The fifth profile, the 
top curve in the figure, corresponds 
to the case of no drug delivery system, 
and has been included for comparison. 
In this case, all of the growth factor 
is free in the gel, and its 
concentration is governed by the usual 
linear diffusion equation (see 
Section \ref{sec:nodelivery}). The 
other four curves correspond to the 
four asymptotic regimes identified in 
the previous Section. We observe for 
these that the growth factor release 
rate becomes slow after a period of 
appproximately a day, and this is 
easily interpreted. The fast initial 
release phase corresponds to the 
rapid out-diffusion of free growth 
factor on a diffusion time scale; 
notice that this period coincides 
(tellingly) with the period over 
which the growth factor releases when 
there is no delivery system. Once 
the free component has substantially 
exited the system, the remaining 
bound fraction releases slowly over 
a long time scale, as previously 
explained. In all cases, the 
numerical results are consistent
with the asymptotic predictions.

In Figure \ref{fg:vary_parameters} (b), 
we display numerical solutions to 
(\ref{eq:reduced_governing_system})
for
$K_{b{\mbox{\tiny P}}},
\eta_{\mbox{\tiny H/G}}\gg 1$, and for
various values of 
$K_{b{\mbox{\tiny H}}}$. These 
parameters correspond to the heparin
being strongly retained by the peptide,
and the initial concentration of
heparin greatly exceeding that of
the growth factor. However, we see
from the figure that this is not 
sufficient to guarantee slow 
release of growth factor. For values 
$K_{b{\mbox{\tiny H}}}=O(1)$, 
which correspond to moderate 
retention of growth factor, the
growth factor will release over a
period of some days. This compares 
unfavourably with the results displayed
in Figure \ref{fg:vary_parameters} 
(a) where both $K_{b{\mbox{\tiny P}}}$ 
and $K_{b{\mbox{\tiny H}}}$ are large 
and we have slow release in all 
cases even though the values for 
$\eta_{\mbox{\tiny H/G}}$ there
are only $O(1)$.


\subsection{Assessing the 
validity of the model
experimentally}
\label{sec:validate_model}

The appropriateness of the 
proposed governing mathematical 
model may be assessed experimentally 
for a particular system by simply 
omitting components from  the 
polymerization mixture when preparing 
the fibrin gels. The governing mathematical 
model may then reduce considerably, making 
it easier to compare its predictions 
with experimental data and enabling 
parameter estimation. We suggest that 
such simpler systems should be 
assessed experimentally as a 
preliminary to consideration of the 
complete release system. We consider 
two such cases, and then make a
brief remark concerning the 
full system.

\subsubsection{No heparin}
\label{sec:nodelivery}

If in the preparation of the fibrin 
matrices, no heparin is added to the
fibrinogen solution, then 
$c_{\mbox{\tiny H}}=c_{\mbox{\tiny GH}}=
c_{\mbox{\tiny HP}}=c_{\mbox{\tiny GHP}}=0$. 
The dimensional form for the governing 
equations reduces to:
\begin{eqnarray}
&& \frac{\partial c_{\mbox{\tiny G}}}{\partial t}
  = D_{\mbox{\tiny G}} 
  \frac{\partial^2 c_{\mbox{\tiny G}}}{\partial x^2}, 
  \label{eq:no_heparin} \\
&& \frac{\partial c_{\mbox{\tiny G}}}{\partial x}=0 
   \mbox{ on }x=0, \hspace{0.2cm}
   c_{\mbox{\tiny G}}=0\mbox{ on }x=L, \hspace{0.2cm}
   c_{\mbox{\tiny G}}=c_{\mbox{\tiny G}}^0 
   \mbox{ at }t=0, \nonumber
\end{eqnarray}
and this is easily solved by separating 
variables (\cite{Crank-1975}) to obtain:
\[
c_{\mbox{\tiny G}}(x,t) = 
\frac{{4c_{\mbox{\tiny G}}^0}}{\pi }
\sum\limits_{n = 1}^\infty  
\frac{{(-1)^{n + 1} }}{{2n - 1}}
\exp \left( {\frac{{ - (2n - 1)^2 \pi ^2 
D_{\mbox{\tiny G}} t}}{4L^2}} \right)
\cos \left(\frac{{(2n - 1)\pi x}}{2L} 
\right).
\]
The total amount of growth factor 
released from the fibrin matrix by 
time $t$ is then given by:
\[
M(t) = L c_{\mbox{\tiny G}}^0 - 
\int\limits_{0}^{L} 
c_{\mbox{\tiny G}}(x,t)dx,
\]
from which it follows that the fraction 
of the available growth factor released 
by time $t$ is: 
\begin{equation}
\frac{M(t)}{M(\infty)}= 
 {1 - \frac{8}{{\pi ^2 }}
 \sum\limits_{n = 1}^\infty  
 \frac{1}{{(2n - 1)^2 }} 
 \exp \left( \frac{{ - (2n - 1)^2 \pi^2 
 D_{\mbox{\tiny G}} t}}{4L^2} \right) }.
\label{eq:fraction_released1}
\end{equation}
This expression contains only one unknown
parameter, $D_{\mbox{\tiny G}}$, and it
predicts that the growth factor should
release on the time scale 
$t=O(L^{2}/D_{\mbox{\tiny G}})$, which
for the matrices described here and
typical growth factors corresponds to a 
period of some days. We suggest a four
day release experiment and at least
three data points per day for the 
fraction of growth factor released. 
This release data may then be 
compared with 
(\ref{eq:fraction_released1}). If 
there is a poor match between the 
experimental and theoretical profiles,
or if one finds that an unreasonable 
value for $D_{\mbox{\tiny G}}$ must be
used to obtain an acceptable fit, then
the appropriateness of the model for 
the delivery system of interest is
called into question. It may be that
some other process not incorporated
in the modelling is significantly
affecting the release behaviour.

\subsubsection{No growth factor and peptide}

If both growth factor and peptide are 
omitted from the polymerization mixture, 
then 
$c_{\mbox{\tiny G}}=c_{\mbox{\tiny P}} = 
c_{\mbox{\tiny GH}} = c_{\mbox{\tiny HP}} = 
c_{\mbox{\tiny GHP}} = 0$. The only 
surviving species in the model is free 
heparin, and its concentration is 
governed by an initial boundary value 
problem identical in structure to 
(\ref{eq:no_heparin}); simply substitute 
$G$ with $H$ in (\ref{eq:no_heparin}) 
and (\ref{eq:fraction_released1}). We 
now suggest a four day release 
experiment for the {\em heparin} with 
at least three data points per day for 
the fraction of heparin released. The 
experimental and theoretical release 
results may then be compared, providing 
a second test of the validity of the 
model. If the correspondence between 
theory and experiment is good, then the 
heparin diffusivity $D_{\mbox{\tiny H}}$ 
is estimated (of course the value 
obtained must be consistent with 
previous estimates; it must have 
the correct order of magnitude).

\subsubsection{The full system}

If the model passes the tests set 
for it in the previous two 
subsections, one may proceed to 
preparing the matrix with all 
components included. If the matrix 
is prepared in accordance with 
the recommendations 
(\ref{eq:full_criteria}), one 
should experimentally observe the 
growth factor release rate become 
slow after a period of a few days 
if the model is valid.


\section{Comparison with 
experimental data 
\label{sec:comparison}}


We now compare the theoretical 
release profiles generated by 
the model with in vitro 
experimental release data. 
For experimental data where 
there is no delivery system or
no heparin, we make the comparison
with the model for all times 
using the analytical expression
(\ref{eq:fraction_released1}).
However, for experimental data
where the full delivery system
is present, we do not attempt to 
compare the model results with 
experimental data in the first 
two days of release since free 
peptide may play a significant 
role in this period, and the 
model does not track the 
concentration of this species. 
In fact, to incorporate the 
effect of free peptide would 
require the inclusion of three 
more species in the model: free 
peptide, free peptide-heparin 
complex, and free 
peptide-heparin-growth factor
complex; see \cite{Wood-2007}. 
This would add three 
reaction-diffusion equations
to the governing mathematical 
model, and would complicate the 
analysis considerably. However, 
after a period of some days, all 
of these species should have 
substantially diffused out of the 
system since there is no mechanism 
to replenish them (the covalently 
bound peptide does not dissociate), 
and the species that then remain 
do form part of the model described 
here. We should say that it is not 
difficult to fit the model results 
with complete release profiles that 
include the first few days, but this 
would require the selection of 
parameters in the model that are 
not compatible with the experimental 
conditions. The numerical solutions
displayed in 
Figure \ref{fg:vary_parameters} (a) 
do have the qualitative character 
of many experimental profiles.

In selecting parameter values for 
the model, we use wherever possible
the values used in the experiments. 
We always use the same heparin to 
growth factor ratio 
$\eta_{\mbox{\tiny H/G}}$ and heparin
to peptide ratio $\eta_{\mbox{\tiny H/P}}$
as used in the experiments. Where estimates 
for the diffusivities can be found, either
in the experimental paper in question, or
elsewhere, we use them. Where a 
value for $D_{\mbox{\tiny G}}$ is 
not available, we estimate it by 
fitting an experimental 
profile for no delivery system to the 
corresponding theoretical release profile 
(\ref{eq:fraction_released1}). For 
$D_{\mbox{\tiny GH}}$, we then select 
a value which has order of magnitude 
10$^{-5}$ cm$^{2}$/min, and which is 
such that $D_{\mbox{\tiny GH}}<
D_{\mbox{\tiny G}},D_{\mbox{\tiny H}}$.
The values for $K_{b{\mbox{\tiny P}}} = 
c_{\mbox{\tiny P}}^{\mbox{\tiny 0}}/
K_{\mbox{\tiny H-P}}^{\mbox{\tiny D}}$
are calculated using the given
values for 
$c_{\mbox{\tiny P}}^{\mbox{\tiny 0}}$
and
$K_{\mbox{\tiny H-P}}^{\mbox{\tiny D}}
=8.67\times 10^{-8}$ M. This is probably
an over-estimate since we are beginning our simulation after day two
and unbound peptide will have been lost,
reducing the value for
$c_{\mbox{\tiny P}}^{\mbox{\tiny 0}}$. 
A similar remark applies to the values of
$c_{\mbox{\tiny G}}^{\mbox{\tiny 0}}$
and 
$c_{\mbox{\tiny H}}^{\mbox{\tiny 0}}$.
However, the values for 
$K_{b{\mbox{\tiny P}}}$ are
of the order of thousands, and 
adjusting them by a factor of 
two or so will have very little 
effect on the resulting profiles. 
The selection of appropriate values 
for 
$K_{b{\mbox{\tiny H}}}=
c_{\mbox{\tiny H}}^{\mbox{\tiny 0}}/
K_{\mbox{\tiny G-H}}^{\mbox{\tiny D}}$ 
is a much more delicate issue though 
because its values range considerably 
in the experiments and the behaviour 
is usually strongly dependent on this 
value. Unfortunately, both numbers 
involved in the calculation of 
$K_{b{\mbox{\tiny H}}}$ are uncertain 
here since 
$c_{\mbox{\tiny H}}^{\mbox{\tiny 0}}$
needs to be reduced as explained above 
and only order of magnitude estimates
are available for
$K_{\mbox{\tiny G-H}}^{\mbox{\tiny D}}$.
Hence, we use $K_{b{\mbox{\tiny H}}}$
as a fitting parameter, but insist that
it has the same order of magnitude 
as $c_{\mbox{\tiny H}}^{\mbox{\tiny 0}}/
K_{\mbox{\tiny G-H}}^{\mbox{\tiny D}}$
where 
$c_{\mbox{\tiny H}}^{\mbox{\tiny 0}}$
is the value given in the experiment
and
$K_{\mbox{\tiny G-H}}^{\mbox{\tiny D}}$
is the order of magnitude estimate for
this dissociation constant.

We have chosen to compare the model
results with experimental data drawn
from three studies:
(a) \cite{Taylor-2004}, 
(b) \cite{Wood-2007}, and 
(c) \cite{Wood-2008}; we shall 
subsequently refer to these as 
(a),(b),(c), and this labelling has 
also been used in Figure \ref{fg:fitting} 
where the comparison between the 
model and the experimental data is 
given. In (a), (b), (c) the growth 
factors used are NT-3, NGF and GDNF, 
respectively. In Figure \ref{fg:fitting}, 
we also display in each case the 
solution (\ref{eq:fraction_released1}) 
with the appropriate value for 
$D_{\mbox{\tiny G}}$, which is the 
theoretical prediction when there is 
no delivery system or no heparin.
In Figure \ref{fg:fitting} (a), we 
display experimental data with
$K_{b{\mbox{\tiny H}}}\sim 1$ and 
$K_{b{\mbox{\tiny H}}}\sim 5$, but we
do not attempt to fit this data other
than to comment that it follows quite
closely the behaviour of the solution
(\ref{eq:fraction_released1}).

We note that the correspondence 
between the model and experimental
data is satisfactory in all cases. 
It is clear that for
$K_{b{\mbox{\tiny H}}}\gg 1$, the 
experimental release rates becomes 
slow after a few days, which is 
consistent with the theoretical 
prediction. For $O(1)$ values of 
$K_{b{\mbox{\tiny H}}}$ 
(Figure \ref{fg:fitting} (a)),
the experimental realease rates
are comparable to that for no 
delivery system. The experimental 
release profiles for data corresponding 
to no delivery system or no heparin 
are adequately described by 
(\ref{eq:fraction_released1}).


\section{Discussion}
	
We summarise our results.

\begin{itemize}

\item We have shown that for conditions 
      of typical interest, the governing 
      mathematical model may be reduced 
      to a system of just two partial 
      differential equations, and that the 
      release behaviour is frequently 
      dominated by the values of two 
      non-dimensional parameters. If the 
      model is valid for a particular system, 
      there will usually be slow passive 
      release of at least a fraction of 
      the growth factor if the fibrin 
      matrices are prepared with the 
      concentration of crosslinked peptide 
      greatly exceeding the dissociation 
      constant of heparin from peptide, and 
      the concentration of heparin greatly 
      exceeding the dissociation constant 
      of growth factor from heparin. It is
      noteworthy that these criteria do not 
      preclude slow release for growth factors 
      that bind heparin with low affinity. 
      We also note the value of
      having reliable estimates for the two 
      dissociation constants in the system. 

\item It is experimentally convenient to
      vary the ratios of heparin to growth
      factor and of heparin to peptide in the 
      polymerisation mixture for the gels 
      to determine optimal conditions for slow
      passive release. However, these ratios
      are not usually the key parameters, and
      where this strategy does result in slow
      release, we have found that it is because
      the binding constants have strayed into 
      the regime referred to in the point 
      immediately above. Our results indicate 
      that the ratios of heparin to growth 
      factor and of heparin to peptide in the 
      polymerisation mixture need neither be 
      large nor small for slow release.      

\item For the first time, theoretical 
      release profiles generated by the 
      model are compared directly with 
      in vitro experimental data. It is 
      found that once the free components 
      have cleared the system, the 
      correspondence between experimental 
      and theoretical results is 
      satisfactory. In particular, our 
      predictions concerning conditions 
      that will give rise to slow passive 
      release are confirmed.

\item It may be possible to partially unpick 
      the system experimentally by simply 
      omitting components in the polymerization 
      mixture for the fibrin gels. For example, 
      if heparin is omitted from the 
      polymerization mixture, the governing 
      mathematical model reduces to a standard 
      linear diffusion equation for the growth 
      factor, and theoretical predictions may 
      then be readily compared with 
      experimental data to help validate the 
      model and estimate a diffusivity; see 
      Section \ref{sec:validate_model}.            

\end{itemize}


\newpage

\vspace{0.5cm}

\noindent {\bf Acknowledgement}

\vspace{0.3cm}

VTNT thanks the Mathematics Applications 
Consortium for Science and Industry (MACSI) 
and Science Foundation Ireland (SFI) for 
their financial support (06/MI/005). MGM 
thanks MACSI for its support and NUI Galway 
for the award of a travel grant.


\renewcommand{\theequation}{A.\arabic{equation}}   
\setcounter{equation}{0}  
\section*{Appendix}  

\noindent {{\em Species concentrations in 
terms of total growth factor and total 
heparin}}

\vspace{0.3cm}

Solving the six algebraic expressions 
(\ref{eq:nondimensional_governing_equations})$_{6}$,
(\ref{eq:nondimensional_totals})$_{1}$,
(\ref{eq:equilibrium_equations})
for the concentrations of the six species
$c_{\mbox{\tiny G}}$, $c_{\mbox{\tiny H}}$,
$c_{\mbox{\tiny GH}}$, $c_{\mbox{\tiny P}}$,
$c_{\mbox{\tiny HP}}$, $c_{\mbox{\tiny GHP}}$
in terms of the total concentration of
growth factor, 
$c_{\mbox{\tiny G}}^{\mbox{\tiny T}}$,
and heparin, 
$c_{\mbox{\tiny H}}^{\mbox{\tiny T}}$, 
gives:
\begin{eqnarray}
&& c_{\mbox{\tiny G}}(
   c_{\mbox{\tiny G}}^{\mbox{\tiny T}},
   c_{\mbox{\tiny H}}^{\mbox{\tiny T}})
 = \frac{ 
   c_{\mbox{\tiny G}}^{\mbox{\tiny T}} -
   \eta_{\mbox{\tiny H/G}} 
   c_{\mbox{\tiny H}}^{\mbox{\tiny T}} - 
   \frac{\eta_{\mbox{\tiny H/G}}}
  {K_{b{\mbox{\tiny H}}}} + 
  \sqrt{\left( 
  c_{\mbox{\tiny G}}^{\mbox{\tiny T}} -
  \eta_{\mbox{\tiny H/G}}  
  c_{\mbox{\tiny H}}^{\mbox{\tiny T}} - 
  \frac{\eta_{\mbox{\tiny H/G}}}
  {K_{b{\mbox{\tiny H}}}} \right)^2 + 
  \frac{4 \eta_{\mbox{\tiny H/G}}
  c_{\mbox{\tiny G}}^{\mbox{\tiny T}}}
  {K_{b{\mbox{\tiny H}}}}} }
  {2},  \nonumber \\
&& \nonumber \\
&& c_{\mbox{\tiny P}}(
    c_{\mbox{\tiny H}}^{\mbox{\tiny T}}) 
 = \frac{1 - 
   \frac{1}{K_{b{\mbox{\tiny P}}}} - 
   \eta_{\mbox{\tiny H/P}}    
   c_{\mbox{\tiny H}}^{\mbox{\tiny T}} +  
   \sqrt{\left( \eta_{\mbox{\tiny H/P}} 
   c_{\mbox{\tiny H}}^{\mbox{\tiny T}} - 
    1 - \frac{1}{K_{b{\mbox{\tiny P}}}} 
   \right)^2 + 
   \frac{4 \eta_{\mbox{\tiny H/P}} 
   c_{\mbox{\tiny H}}^{\mbox{\tiny T}}}
   {K_{b{\mbox{\tiny P}}}}} }{2}, 
   \nonumber \\
&& \nonumber \\
&& c_{\mbox{\tiny GH}}(
    c_{\mbox{\tiny G}}^{\mbox{\tiny T}},
    c_{\mbox{\tiny H}}^{\mbox{\tiny T}})
 = \frac{ 
    \frac{\eta_{\mbox{\tiny H/G}}}
    {\eta_{\mbox{\tiny H/P}}} 
    (1-c_{\mbox{\tiny P}}
    (c_{\mbox{\tiny H}}^{\mbox{\tiny T}}))
    \left\{ \frac{1}{K_{b{\mbox{\tiny P}}}
    c_{\mbox{\tiny P}}(
   c_{\mbox{\tiny H}}^{\mbox{\tiny T}})} 
   +\frac{1}{1+\frac{K_{b{\mbox{\tiny H}}}\theta
   c_{\mbox{\tiny G}}(
   c_{\mbox{\tiny G}}^{\mbox{\tiny T}},
   c_{\mbox{\tiny H}}^{\mbox{\tiny T}})}
   {\eta_{\mbox{\tiny H/G}}(1+\theta)}  
   }\right\} -\frac{\eta_{\mbox{\tiny H/G}}
   (c_{\mbox{\tiny G}}^{\mbox{\tiny T}}-
   c_{\mbox{\tiny G}}(
    c_{\mbox{\tiny G}}^{\mbox{\tiny T}},
    c_{\mbox{\tiny H}}^{\mbox{\tiny T}}))}
    {K_{b{\mbox{\tiny H}}}c_{\mbox{\tiny G}}(
    c_{\mbox{\tiny G}}^{\mbox{\tiny T}},
    c_{\mbox{\tiny H}}^{\mbox{\tiny T}}))}}                 
 {1+\frac{\frac{K_{b{\mbox{\tiny P}}}
  c_{\mbox{\tiny P}}(
   c_{\mbox{\tiny H}}^{\mbox{\tiny T}})}
   {1+\theta}}{1+\frac{K_{b{\mbox{\tiny H}}}\theta
   c_{\mbox{\tiny G}}(
   c_{\mbox{\tiny G}}^{\mbox{\tiny T}},
   c_{\mbox{\tiny H}}^{\mbox{\tiny T}})}
   {\eta_{\mbox{\tiny H/G}}(1+\theta)}}},  
   \nonumber \\
&& \label{eq:species_in_terms_of_totals} \\
&& c_{\mbox{\tiny GHP}}(
     c_{\mbox{\tiny G}}^{\mbox{\tiny T}},
     c_{\mbox{\tiny H}}^{\mbox{\tiny T}})
 = c_{\mbox{\tiny G}}^{\mbox{\tiny T}}
   - c_{\mbox{\tiny G}}(
   c_{\mbox{\tiny G}}^{\mbox{\tiny T}},
   c_{\mbox{\tiny H}}^{\mbox{\tiny T}}) 
   - c_{\mbox{\tiny GH}}(
    c_{\mbox{\tiny G}}^{\mbox{\tiny T}},
    c_{\mbox{\tiny H}}^{\mbox{\tiny T}}), 
     \nonumber \\
&& \nonumber \\
&& c_{\mbox{\tiny HP}}(
     c_{\mbox{\tiny G}}^{\mbox{\tiny T}},
     c_{\mbox{\tiny H}}^{\mbox{\tiny T}})
 =  \frac{ \frac{1}{\eta_{\mbox{\tiny H/P}}}
    (1-c_{\mbox{\tiny P}}(
    c_{\mbox{\tiny H}}^{\mbox{\tiny T}}))
    -\frac{  K_{b{\mbox{\tiny P}}} 
    c_{\mbox{\tiny GH}}(
    c_{\mbox{\tiny G}}^{\mbox{\tiny T}},
    c_{\mbox{\tiny H}}^{\mbox{\tiny T}})
    c_{\mbox{\tiny P}}(
    c_{\mbox{\tiny H}}^{\mbox{\tiny T}}) 
    }{\eta_{\mbox{\tiny H/G}}(1+\theta)}}
    { 1 + \frac{  K_{b{\mbox{\tiny H}}} 
    c_{\mbox{\tiny G}}(
    c_{\mbox{\tiny G}}^{\mbox{\tiny T}},
    c_{\mbox{\tiny H}}^{\mbox{\tiny T}})
    \theta }
    {\eta_{\mbox{\tiny H/G}}(1+\theta)} },
 \nonumber \\
&&  \nonumber \\
&& c_{\mbox{\tiny H}}(
    c_{\mbox{\tiny G}}^{\mbox{\tiny T}},
    c_{\mbox{\tiny H}}^{\mbox{\tiny T}})
 = c_{\mbox{\tiny H}}^{\mbox{\tiny T}}
   - c_{\mbox{\tiny HP}}(
   c_{\mbox{\tiny G}}^{\mbox{\tiny T}},
   c_{\mbox{\tiny H}}^{\mbox{\tiny T}})
   - (c_{\mbox{\tiny GH}}(
    c_{\mbox{\tiny G}}^{\mbox{\tiny T}},
    c_{\mbox{\tiny H}}^{\mbox{\tiny T}})
   + c_{\mbox{\tiny GHP}}(
    c_{\mbox{\tiny G}}^{\mbox{\tiny T}},
    c_{\mbox{\tiny H}}^{\mbox{\tiny T}}))
    /\eta_{\mbox{\tiny H/G}}. 
    \nonumber
\end{eqnarray}
Hence, in the model considered  here, it 
is sufficient to solve for 
$c_{\mbox{\tiny G}}^{\mbox{\tiny T}}$
and
$c_{\mbox{\tiny H}}^{\mbox{\tiny T}}$.


\begin{table}[!b] 
\tblcaptionnotes{The data used to generate 
the theoretical curves in Figure \ref{fg:fitting}.
Unmarked data is taken from the paper referred to
in its column. The markings on the remaining
data are explained below the table.}
{\mbox{\tabcolsep=4pt \begin{tabular}{@{}lcccl@{}}
\tblhead{ {} & Figure \ref{fg:fitting} (a) & Figure \ref{fg:fitting} (b)  & Figure \ref{fg:fitting} (c)\\ [3pt]
Symbol       &  NT-3 (\cite{Taylor-2004})   &  NGF (\cite{Wood-2007})   &  GDNF (\cite{Wood-2008})    & Units \\[-9.5pt]}\\[-9.5pt]
$D_{\mbox{\tiny G}}$  & ${2.5 \times 10^{-5}}^{(a)}$        & $9.7 \times 10^{-5}$      &  ${2.0 \times 10^{-5}}^{(a)}$       & cm$^2$/min     \\[2pt]
$D_{\mbox{\tiny H}}$  & ${3.0 \times 10^{-5}}^{(b)}$        & $9.1 \times 10^{-5}$      &  ${3.0  \times 10^{-5}}^{(b)}$      & cm$^2$/min     \\[2pt]
$D_{\mbox{\tiny GH}}$ & ${1.0 \times 10^{-5}}^{(c)}$        & $7.5 \times 10^{-5}$      &  ${1.5 \times 10^{-5}}^{(a)}$       & cm$^2$/min     \\[2pt]
$K_{\mbox{\tiny G-H}}^{\mbox{\tiny D}}$  & ${1.5 \times 10^{-7}}^{(a)}$        & ${0.33 \times 10^{-7}}^{(a)}$     &  ${1.25 \times 10^{-7}}^{(a)}$            & M     \\[2pt]
$K_{\mbox{\tiny H-P}}^{\mbox{\tiny D}}$  & ${8.67 \times 10^{-8}}^{(d)}$        & ${8.67 \times 10^{-8}}^{(d)}$     &  ${8.67 \times 10^{-8}}^{(d)}$      & M     \\[2pt]
$c_{\mbox{\tiny P}}^{\mbox{\tiny 0}}$  & $2.3 \times 10^{-4}$        & $2.5 \times 10^{-4}$     &  $2.5 \times 10^{-4}$      & M     \\[3pt]
$c_{\mbox{\tiny G}}^{\mbox{\tiny 0}}$  & $7.4 \times 10^{-9}$        & $7.4 \times 10^{-9}$     &  $3.1 \times 10^{-9}$      & M   \lastline  
\end{tabular}
\label{table:fitting}
}}
{$^{(a)}$ Estimated from data in \cite{Taylor-2004, Wood-2007, Wood-2008}.\\
$^{(b)}$ \cite{Gaigalas-1995}.\\
$^{(c)}$ \cite{Saltzman-1994}.\\
$^{(d)}$ ${\mbox{\tiny K}}_{\mbox{\tiny R}}= 78$ min$^{-1}$  
(\cite{Olson-1981}) and 
${\mbox{\tiny K}}_{\mbox{\tiny F}}= 9.0 \times 10^8$ M$^{-1}$min$^{-1}$  (\cite{Tyler-Cross-1994, Tyler-Cross-1996, Kridel-1996}).}
\vspace*{-2pt}
\end{table}

\end{document}